\documentstyle[preprint,aps]{revtex}
 \tightenlines

\begin{document}

\title{Attractive and Repulsive Gravity\footnote{gr-qc/0001011, 
December 31, 1999}}

\author{\normalsize{Philip D. Mannheim} \\
\normalsize{Department of Physics,
University of Connecticut, Storrs, CT 06269} \\
\normalsize{mannheim@uconnvm.uconn.edu} \\}

\maketitle

\begin{abstract}
We discuss the circumstances under which gravity might be repulsive rather 
than attractive. In particular we show why our standard solar system distance 
scale gravitational intuition need not be a reliable guide to the behavior of 
gravitational phenomena on altogether larger distance scales such as 
cosmological, and argue that in fact gravity actually gets to act
repulsively on such distance scales. With such repulsion a variety of 
current cosmological problems (the flatness, horizon, dark matter, universe 
age, cosmic acceleration and cosmological constant problems) are then all 
naturally resolved.

\end{abstract}

\section{Introduction}

Few observational facts appear to be as well established in physics as the 
attractive nature of gravity. However, despite this, recent cosmological 
observations \cite{Riess1998,Perlmutter1998} have raised the possibility that 
under certain circumstances gravity might actually contain an effective 
repulsive component, to thus invite consideration of the degree to which, and 
of the specific set of conditions under which, gravity actually need be  
strictly attractive in the first place. In fact, familiar as attractive 
gravity is, its actual attractiveness stems from the a priori assumption that 
Newton's constant $G$ be chosen (purely by hand) to actually be positive. Then 
with this positive $G$ being treated as a fundamental input parameter by both 
Newtonian gravity and its relativistic Einstein gravity generalization (and 
even by their quantum-mechanical string theory generalization as well for that 
matter), the universal attractive nature of gravity is then posited on all 
distance scales and for all possible gravitational field strengths. As such, 
this actually constitutes a quite severe extrapolation of observationally 
established gravitational information from the kinematic solar system distance 
scale weak gravity regime where it was expressly obtained in the first place. 
And indeed, when the standard Newton-Einstein gravitational theory is actually
extended beyond its solar system origins, notwithstanding the successes that 
are then encountered, nonetheless, disturbingly many difficulties are also  
encountered, in essentially every single such type of extrapolation that is in
in fact made. Thus its extrapolation to larger distances such as galactic leads 
to the need for as yet unestablished galactic dark matter, its extrapolation to 
strong gravity leads to singularities and the development of event horizons and 
trapped surfaces in the fabric of spacetime, its extrapolation to the high 
temperature early universe leads to a cosmology with fine tuning and 
cosmological dark matter problems as well as to the notorious cosmological 
constant problem, and its extrapolation to large quantum field theoretic 
momenta far off the mass shell leads to uncontrollable renormalization 
infinity problems. Now while all of these issues may ultimately be resolved in 
favor of the standard theory, it is important to emphasize that 
all of them essentially arise from using just the first few measured weak, 
perturbative terms in a series (such as the first few terms in the 
Schwarzschild metric, the only such terms in that metric which have so far 
been observationally tested in fact) to try to guess the rest of the series. 
With there thus being many possible extrapolations of the standard 
Schwarzschild weak gravity solar system wisdom, extrapolations which can be 
just as covariant as the standard one, there are thus many possible departures 
from the standard gravitational intuition when gravity is extended to 
altogether different conditions, and in this paper we shall explore the 
possibility that, even as it acts attractively on solar system distance 
scales, gravity nonetheless gets to act repulsively on the much larger one 
associated with cosmology. 

\section{How the Standard Intuition Came About}

In developing a fundamental theory of planetary motion Newton found that an
inverse square force law with a universal gravitational constant $G$ not 
merely accounted for the Keplerian elliptical orbital motions of the planets 
around the sun, but equally, it also described the gravitational motions of 
objects near the surface of the earth, i.e. it described gravity not only on 
solar system distance scales but also on altogether smaller terrestrial 
distance scales as well. In this way Newton's law of gravity thereby acquired 
a universal character causing it to come to be regarded as a universal law 
which was then to also be valid on altogether larger distance scales as
well even though it had not in fact been tested on them. However, with  
the advent of galactic astronomy eventually then permitting such testing on 
these altogether larger distance scales, it was actually found (see e.g. 
\cite{Trimble1987,Binney1998} for recent reviews) that the detailed orbits of 
stars and galactic gas in galaxies did not in fact conform to the ones expected 
on the basis of the known luminous matter content of the galaxies. In fact long 
ago Zwicky \cite{Zwicky1933} had already noted an analogous problem in (presumed 
virialized) clusters of galaxies, with the discrepancy he found between the 
measured mean kinetic energy of the visible galaxies within the cluster and the 
mean gravitational potential energy generated by those selfsame galaxies 
leading him to conclude that the continuing applicability of Newton's law of 
gravity entailed that there would have to be far more matter in the 
cluster than he was able to detect. Then, once it became clear that 
discrepancies such as these were in fact common and even widespread on these 
large distance scales, a dark matter paradigm was then adopted, a paradigm in 
which any detected discrepancy between the known luminous matter Newtonian 
expectation and observation was to then be accounted for by whatever amount of 
non-luminous matter would then (i.e. only after the fact) specifically be 
required. Moreover, with there not as yet being a single application or test of 
Newtonian gravity on these large distance scales (in either galaxies or clusters 
of galaxies) which does not involve an appeal to this as 
yet unestablished and still poorly understood dark matter, we thus see the 
complete circularity of the chain of reasoning which leads to dark matter in 
the first place, since one assumes the continuing validity of Newton's law of 
gravity and then posits the presence of just the appropriate amount of dark 
matter needed in order to maintain the validity of the presupposed Newtonian 
law. Apart from not being a particularly satisfying prescription, this 
galactic dark matter paradigm does yet even qualify as being a falsifiable 
theory (the sine qua non of physical theory) since it has not yet been brought 
to the point where it can actually make definitive and expressly falsifiable 
galactic predictions, i.e. given a detected amount of luminous matter in a 
given galaxy, dark matter theory should be able to predict both the amount and 
spatial distribution of the dark matter in the galaxy in advance of any 
measurement of a galactic rotation curve. Moreover, even more disturbing than 
this is the concomitant need of having to now retroactively explain why dark 
matter is not also required in order to fit solar system planetary orbits, 
i.e. the need to explain why luminous matter alone should in fact be 
capable of providing a complete accounting of solar system dynamics in the 
first place. A model in which dark matter is there when needed and not when 
not hardly qualifies as being a model at all, with its most conservative 
characterization actually being that in fact dark matter is nothing more than 
a phenomenological parameterization of the detected departure of the luminous 
Newtonian expectation from observation. This then is the galactic dark 
matter problem, and while it may eventually be resolved by the actual direct 
detection of dark matter, nonetheless it could equally well be signaling a 
failure of the standard Newtonian wisdom and intuition on these very large 
distance scales. 

While some Newtonian wisdom was in fact supplanted by the subsequent 
development of Einstein relativity, it is curious that its development in fact
actually served to reinforce the particular Newtonian intuition that we  
described above. Specifically, even while the general relativistic curved 
spacetime Einstein gravitational theory replaced the strictly non-relativistic 
Newtonian gravitational one, nonetheless the Einstein theory still recovered 
the Newtonian theory in the non-relativistic limit, even as it prescribed 
relativistic corrections to it. The actual observational confirmation of these
relativistic corrections not merely served to establish the validity of
Einstein gravity, it also served to reinforce the validity of Newtonian gravity 
whenever the non-relativistic limit could appropriately be taken. However, 
since these relativistic corrections were themselves only established on solar 
system distance scales (cf. the first few terms in a perturbative solar system 
Schwarzschild metric expansion or the first few perturbative terms in the 
metric of a binary pulsar, a similarly sized such system), the extrapolation of 
Einstein gravity (and of its universal character) to galactic distance scales 
and beyond was then no more secure than had been the extrapolation of Newtonian 
gravity to those same distance scales (with the Einstein equations thus only 
being as secure as dark matter). Thus again a new intuition was acquired, one 
in which the very presence of Newton's constant as an a priori fundamental 
coupling constant in the Einstein-Hilbert action then endowed $G$ with an even 
more fundamental and universal status than it had actually previously 
possessed. Moreover, in giving Newton's constant such a status, a dichotomy is 
immediately set up between gravity and the standard $SU(2) \times U(1)$ 
electroweak theory, a theory where another dimensionful phenomenological 
parameter, Fermi's constant $G_F$, is not in fact elevated into a fundamental 
parameter at all, a theory which can then be made completely renormalizable 
precisely because this is not in fact done, with $G_F$ itself then emerging 
solely as an effective parameter which is only of relevance at low energies.

That Einstein gravity was taken to be universal was hardly surprising given 
its very general geometric character. However, it is important to distinguish 
between the geometric and dynamical aspects of the Einstein theory, and even 
while these two issues are logically distinct, nonetheless the Einstein theory 
is ordinarily regarded as being one integral and indivisible package. 
However, the standard theory both geometrizes gravity by identifying the 
spacetime metric $g^{\mu \nu}$ as the gravitational field, and then determines 
its dynamics by imposing the second order Einstein equations
\begin{equation}R^{\mu\nu}-
g^{\mu\nu}R^\alpha_{\phantom{\alpha}\alpha}/2 =-8\pi GT^{\mu\nu}
\label{1}
\end{equation}
where $R^{\mu \nu}$ is the Ricci tensor associated with the geometry and 
$T^{\mu \nu}$ is the energy-momentum tensor of its gravitational source. 
However, even while insisting that the metric is to describe the gravitational 
field, nonetheless, without giving up general covariance, its dynamical 
equations of motions could still depart from the above second order Einstein 
equations, and would in fact readily do so if the gravitational 
equations were to be obtained from the variation of some equally covariant 
general coordinate scalar action other than the standard Einstein-Hilbert one 
($I_{EH}=-\int d^4x(-g)^{1/2}R^\alpha_{\phantom{\alpha}\alpha}/16\pi G$)
which is ordinarily used. (In fact the fully covariant string gravitational 
theory, for instance, replaces the Einstein equations by a set of equations
which contains an entire, infinite series of derivatives of the Riemann 
tensor.) One thus has to distinguish between the fact of curvature (viz. 
geometry) and the amount of curvature (viz. dynamics), and in particular one 
has to go over both the successes and the problems of Einstein gravity to 
ascertain which are due to geometry and which to dynamics, and should it turn 
out that the successes are predominantly due to geometry while the problems 
arise from a particular assumed dynamics, we would then be able to identify the 
extrapolation of the standard gravity equations of motion beyond their solar 
system origin as the root cause of the problems that standard theory currently 
encounters, while not at the same time needing to give up the underlying 
geometrical picture.

Indeed, the very cornerstone of standard gravity, viz. the equivalence
principle, is completely geometrical and has nothing to do with dynamics at
all. Rather it is a statement about the geometric nature of geodesics, with
particles which follow such geodesics then having to uniquely couple to an 
external gravitational field according to
\begin{equation}
{d^2x^{\lambda} \over d\tau^2}
+\Gamma^{\lambda}_{\mu \nu}
{dx^{\mu} \over d\tau}{dx^{\nu } \over d\tau}=0
\label{2}
\end{equation}
regardless of the particular form of the equation of motion obeyed by the 
external gravitational field itself. Thus, even while the magnitudes of the 
Christoffel symbols are of course sensitive to the dynamics associated with the 
background gravitational field, nonetheless, their very  presence in the geodesic 
equations to begin with is strictly geometric. Gravitational bending, lensing, 
redshifting and time delaying of light, modifications to Newtonian planetary 
orbits, and the decay 
of the orbit of a binary pulsar (a consequence of the retarded nature of the 
gravitational radiation reaction of each of the two stars in the binary on the 
other due to the finite limiting velocity with which gravitational information is 
communicated) will thus all be found to occur in any strictly covariant metric 
theory of gravity, with the dynamical equations only controlling the magnitude 
of these effects but not the fact of their existence.\footnote{In passing it is
important to note that there is actually a hidden assumption in using geodesic 
equations to describe gravitational phenomena, viz. the assumption that real
particles can be treated as classical geodesic test particles in the first place. 
Now while the geodesic assumption is immediately valid for the massless rays of 
the eikonal approximation to wave theory (since the rays are already geodesic in 
flat spacetime, with a covariantizing of their motion then making them geodesic 
in a background gravitational field as well), the situation regarding material 
particles is not at all as straightforward, with there actually being no 
justification for associating real particles (i.e. the excitations of the quantum 
fields of elementary particle physics) with the classical test particle action 
$I_T=-mc \int d \tau$, even though the variation of this action would enforce 
geodesic motion. It is thus of interest to note, that in a recent study 
\cite{Mannheim2000,Mannheim1998a} of the extension of the equivalence
principle to the  propagation of quantum-mechanical matter waves in a
background classical  gravitational field (with a curved space Schrodinger
equation explicitly being  shown to be equivalent to an accelerating
coordinate frame flat space one), it  was shown that it is only because there
is such a quantum  extension that the equivalence principle actually gets to
hold for real classical  particles at all. Thus it was shown that gravity,
itself a field theory, couples  first and foremost to fields rather than to
particles (i.e. first and foremost to  wavelength rather than to mass - with
the eikonal rays associated with the  minimally coupled scalar field wave
equation $S^{;\mu}_{\phantom{;\mu};\mu}-(mc/\hbar)^2S=0$ indeed being
geodesic),  and then, only upon second quantization ($\lambda=\hbar/mv$) of
those  rays, to mass. It is thus only because of quantum mechanics 
that massive classical particles get to be geodesic at all.}

Moreover, as regards the uniqueness of any possible underlying dynamics, it was 
noted by Eddington \cite{Eddington1922} in the very early days of relativity 
that the familiar standard gravity exterior $R^{\mu \nu}=0$ Schwarzschild 
solution (the one used in the standard solar system relativistic tests) is just 
as equally a solution to higher derivative gravitational theories as well, 
since the vanishing of the Ricci tensor entails the vanishing of its 
derivatives as well, with solar system tests thus not in fact being able to 
definitively exclude gravitational actions other than the Einstein-Hilbert one 
after all. Moreover, since such higher derivative theories turn out to then 
have different continuations to larger distances \cite{Mannheim1994}, the 
possibility then emerges that the need for galactic dark matter is only an 
artifact of using the  Einstein gravity continuation. And should that be the 
case, the further continuation to cosmology (a regime which is to a good 
degree controlled by the high symmetry of geometries such as Robertson-Walker 
and de Sitter rather than by the structure of the explicit dynamical evolution 
equations themselves) would then potentially become unreliable too. 

In order to illustrate the above remarks in an explicit example it is 
convenient to consider a particular alternate gravitational theory, viz. 
conformal gravity, a general coordinate invariant pure metric theory of 
gravity which possesses an explicit additional and highly restrictive symmetry 
(invariance of the geometry under any and all local conformal stretchings 
$g_{\mu \nu}(x)\rightarrow \Omega^2(x) g_{\mu \nu}(x)$) as well. In fact, so 
restrictive is this symmetry that it allows only one unique gravitational 
action (an action which is to thus replace the standard $I_{EH}$), viz.
\begin{equation} 
I_W=-\alpha_g \int d^4x (-g)^{1/2} C_{\lambda\mu\nu\kappa} 
C^{\lambda\mu\nu\kappa} 
\label{3}
\end{equation}
where $C^{\lambda\mu\nu\kappa}$ is the conformal Weyl tensor and where the  
gravitational coupling constant $\alpha_g$ introduced here is a universal  
dimensionless one, to thereby endow gravity with a structure similar 
to that found in the  electroweak interaction case described above, with such 
a gravitational theory actually then being power counting renormalizable. For
this theory variation of the action leads to the equations of motion
\cite{Dewitt1965}
\begin{equation} 
(-g)^{-1/2}\delta I_W / 
\delta g_{\mu \nu}=-2\alpha_g W^{\mu \nu}=-T^{\mu\nu}/2
\label{4}
\end{equation}
where $W^{\mu \nu}$ is given by   
\begin{eqnarray}
 W^{\mu \nu}=g^{\mu\nu}(R^{\alpha}_{\phantom{\alpha}\alpha})   
^{;\beta} _{\phantom{;\beta};\beta}/2                                             
+ R^{\mu\nu;\beta}_{\phantom{\mu\nu;\beta};\beta}                     
 -R^{\mu\beta;\nu}_{\phantom{\mu\beta;\nu};\beta}                        
-R^{\nu \beta;\mu}_{\phantom{\nu \beta;\mu};\beta}                          
 - 2R^{\mu\beta}R^{\nu}_{\phantom{\nu}\beta}                                    
+g^{\mu\nu}R_{\alpha\beta}R^{\alpha\beta}/2 
\nonumber \\
 -2g^{\mu\nu}(R^{\alpha}_{\phantom{\alpha}\alpha})          
^{;\beta}_{\phantom{;\beta};\beta}/3                                              
+2(R^{\alpha}_{\phantom{\alpha}\alpha})^{;\mu;\nu}/3                             
+2 R^{\alpha}_{\phantom{\alpha}\alpha} R^{\mu\nu}/3                               
-g^{\mu\nu}(R^{\alpha}_{\phantom{\alpha}\alpha})^2/6,                   
\label{5}
\end{eqnarray}
so that we can immediately confirm that the Schwarzschild $R^{\mu \nu}=0$ 
solution is indeed an exterior solution to the theory just as Eddington had 
warned us. Standard gravity is thus seen to be only sufficient to give the 
standard Schwarzschild phenomenology but not at all necessary, with it thus 
indeed being possible to bypass the Einstein-Hilbert action altogether as far 
as low energy phenomena are concerned. 

Further insight into the structure of this alternate gravity theory is 
obtained by noting that for a static, spherically symmetric source such as a 
star, the conformal symmetry allows one \cite{Mannheim1989} to set  
$g_{rr}=-1/g_{00}$ without any loss of generality, with the field equations of 
Eq. (\ref{4}) then being found \cite{Mannheim1994} to reduce (without any 
approximation at all, i.e. for strong and weak gravitational fields alike) to
\begin{equation}    
\nabla ^4 g_{00}=3(T^0_{\phantom{0} 0} -T^r_{\phantom{r} r})/4\alpha_g 
g_{00}\equiv -f(r),
\label{6}
\end{equation}
i.e. to reduce to a fourth order Poisson equation rather than to the second 
order one familiar in the standard theory. The general solution to Eq. (\ref{6}) 
exterior to a star of radius $R$ is then given by 
\cite{Mannheim1994,Mannheim1994a}    
\begin{equation}
-g_{00}(r>R)= 1-2\beta^*/r+\gamma^* r
\label{7}
\end{equation}
with the coefficients being given by 
\begin{equation} 
\beta^{*}=\int_0^R dr f(r) r^4/12,~~  
\gamma^{*}=-\int_0^R dr f(r) r^2/2,
\label{8}
\end{equation}
i.e. by two different moments of the source. We thus see that the Newtonian 
potential need not be associated with either the second order Einstein equations 
or with their non-relativistic second order Poisson equation limit, and that the 
standard theory is thus only sufficient to give Newton but not at all necessary. 
(In order to show the lack of necessity of the standard theory it is sufficient 
to construct just one alternative.) The Newtonian potential can thus just as 
readily be generated in higher order gravitational theories as well, theories 
which can then have a very different behavior on altogether larger distance 
scales. And indeed, through the use of the linear $\gamma^{*} r$ potential term, 
a term which actual dominates over Newton at large enough distances, conformal 
gravity was actually found capable \cite{Mannheim1997,Mannheim1996} of providing 
for a complete accounting of galactic rotation curves without the need to invoke 
dark matter at all. (In fact, the value for the coefficient $\gamma^{*}$ required 
by galactic data then entailed the numerical irrelevance of the $\gamma^{*} r$ 
term on the much smaller solar system distance scales, to thus yield a solution 
of the galactic rotation curve problem which naturally leaves standard solar 
system physics intact.)  However, regardless of the specific merits of any 
particular alternate gravitational theory such as conformal gravity itself, our 
analysis here does serve to underscore the risks inherent in extrapolating solar 
system wisdom beyond the confines of the solar system, with the need for galactic 
dark matter perhaps being symptomatic of the lack of applicability of one 
particular such extrapolation.

With regard to this conformal gravity alternative, we note further, that in it, 
with the coefficient $\beta^{*}$ of a star being given as the energy-momentum 
tensor moment integral of Eq. (\ref{8}), an identification of this same moment 
integral in the case of a single proton or neutron with one half of the 
$2\beta_p$ Schwarzschild radius of a single nucleon, then enables us to identify 
the stellar coefficient $\beta^{*}$ in conformal gravity as $N \beta_p$ for a 
weak gravity star containing $N$ such nucleons. This relation is completely 
identical to the one obtained by solving the standard theory second order Poisson 
equation for a weak gravity source composed of $N$ fundamental elementary 
particle sources each contributing a potential $\beta_p/r$, to thus enable us to 
recover the familiar extensive property of the Newtonian potential of weak 
gravity bulk matter, while also seeing that it need not be tied exclusively to 
the second order theory. Moreover, as far as gravitational sources are concerned, 
the coefficients of their $1/r$ potentials are actually radii, specifically their 
gravitational Schwarzschild radii, so that an identification of $\beta_p$ with 
$Gm_p/c^2$ where $m_p$ is the mass of a proton (to thus define $G$ once and for 
all as $c^2\beta_p/m_p$, a purely microscopic quantity\footnote{In a conformal 
invariant theory dimensionful microscopic parameters such as $\beta_p$ and $m_p$ 
can only be explicitly generated when the conformal symmetry is spontaneously 
broken, with their values then being fixed by the details of the symmetry 
breaking mechanism. However, regardless of any specific dynamics that may be 
needed to explicitly do this, the dependence of the solution of Eq. (\ref{7}) on 
the radial coordinate $r$ is already the most general allowable one possible.}) 
then entails that for a weak gravity bulk matter star $\beta^{*}$ is given as 
$GM^{*}/c^2$ where $M^{*}=Nm_p$ is the mass of the star. We thus see (i) that the 
universality of $G$ need not be tied to the second order Poisson equation, (ii) 
that, just like Boltzmann's constant, Newton's constant $G$ need not itself be 
fundamental (only the product $MG/c^2$ is ever measurable gravitationally and 
never $G$ itself - just as only the product $kT$ is measurable in statistical 
mechanics), and (iii) that $G$ need not have any applicability at all in the 
strong gravity limit where the energy-momentum tensor moment integrals no longer 
scale linearly with the number of fundamental sources. Thus not only might our 
standard gravitational intuition not necessarily be generalizable to large 
distance scales, it might also not be generalizable to strong gravity 
either.\footnote{In fact, for an appropriate gravitational self-energy 
contribution, the moment integrals of Eq. (\ref{8}) might even have differing 
signs in the weak and strong gravity limits.} 

\section{When an Attractive Potential is Repulsive}

Even though it is generally thought that the attractive or repulsive nature of a
potential is determined once and for all by its overall sign, in this section we 
show that this turns out to not in fact necessarily always be the case, with 
this then being another piece of the standard intuition which would appear to 
require reappraisal. To explicitly illustrate this specific point it is 
convenient to consider the geometry near the surface of a static, spherically 
symmetric gravitational source of radius $R$ and mass $M$ with exterior 
geometrical line element of the form  
\begin{equation}
d\tau^2=B(r)c^2dt^2-dr^2/B(r)-r^2d\Omega
\label{8a}
\end{equation}
where the metric coefficient $B(r)$ is given not just by the usual Schwarzschild 
form, but rather by the more general 
\begin{equation}
B(r)=1-2MG/c^2r+\gamma r
\label{8b}
\end{equation}
form we introduced earlier. If we erect a Cartesian coordinate system 
$x=r$sin$\theta$cos$\phi$, $y=r$sin$\theta$sin$\phi$, $z=r$cos$\theta - R$ at the 
surface of the source, then, with $z$ being normal to the surface, to lowest 
order in $x/R,~y/R,~z/R,~MG/c^2R~(=gR/c^2)$ and $\gamma R$ the line element is 
then found \cite{Mannheim2000} to take the form
\begin{equation}
d\tau^2=[1-a(z)]c^2dt^2-dx^2-dy^2-[1+a(z)]dz^2-b(xdx+ydy)dz
\label{9}
\end{equation}
where $a(z)=2g(R-z)/c^2 -\gamma (R+z)$ and $b=4g/c^2-2\gamma$. For trajectories 
for which the initial velocity $v$ is in the horizontal $x$ direction, the 
geodesic equations associated with Eq. (\ref{9}) take the form 
\cite{Mannheim2000} (the dot denotes differentiation with respect to the proper 
time $\tau$ for massive particles or with respect to an affine parameter for 
massless ones)
\begin{equation}
\ddot{t}=0,~ \ddot{x}=0,~ \ddot{y}=0,~
\ddot{z}+2(g-\gamma c^2/2)v^2/c^2+g+\gamma c^2/2=0. 
\label{10}
\end{equation}
Thus in the non-relativistic $v=0$ case the motion is described by
\begin{equation}
\ddot{z}+g+\gamma c^2/2=0, 
\label{11}
\end{equation}
while in the relativistic $v=c$ case it is described by 
\begin{equation}
\ddot{z}+3g-\gamma c^2/2=0 
\label{12}
\end{equation}
instead. Thus we see that the effect of the $\gamma$ term is actually opposite in 
these two limits, with positive $\gamma$ leading to attractive bending for slow 
moving particles but to repulsive deflection for fast moving ones.\footnote{This 
deflection of light away from a source also holds for the exact $B(r)=1
+\gamma r$ geodesic as well \cite{Walker1994,Edery1998}.} Thus the fact that a 
potential may be attractive for non-relativistic motions does not in and of 
itself mean that it must therefore also be attractive for light, with the 
$v^2/c^2$ type terms not only modifying the magnitude of the effect of gravity 
(something already the case even in the standard theory where $\ddot{z}+g(1
+2v^2/c^2)=0$), but even being able to modify the sign of the effect as well. 
Thus in general we see that even after appropriately fixing the sign of the 
coefficient of a gravitational potential term once and for all, such a potential 
need not always lead to attraction, with a potential which would ordinarily be 
considered to be attractive (as defined by non-relativistic binding) still being 
able to lead to repulsion in other kinematic regimes. Caution thus needs to be 
exercised before one can conclude that attraction in one kinematic regime entails 
attraction in all others as well.

As regards the discussion of the metric given in Eq. (\ref{9}) some further
comment is in order. Specifically, noting that the transformation
\begin{eqnarray}
t^{\prime}=t[1-g(R-z)/c^2+\gamma (R+z)/2],~x^{\prime}=x,~y^{\prime}=y,~
\nonumber \\ 
z^{\prime}=z[1+g(2R-z)/2c^2-\gamma (2R+z)/4]
+(g+\gamma c^2/2)t^2/2+(g/c^2-\gamma /2)(x^2+y^2)
\label{13}
\end{eqnarray}
brings the metric of Eq. (\ref{9}) to the flat coordinate form 
$d\tau^2=c^2dt^{\prime 2}-dx^{\prime 2}-dy^{\prime 2}-dz^{\prime 2}$, we see
that the unprimed system origin obeys 
\begin{equation}
z^{\prime}=(g+\gamma c^2/2)t^{\prime 2}/2
\label{14}
\end{equation}
in the primed Cartesian coordinate system. We thus provide a direct demonstration 
of the equivalence principle with gravity indeed being found to act the same way 
as an acceleration in flat spacetime, and with the equivalence principle indeed 
being seen to have validity beyond the standard second order Einstein theory. 

Now while we have just seen that the weak gravity metric of Eq. (\ref{9}) is 
equivalent to an acceleration in flat spacetime, this result is initially 
somewhat puzzling since the full starting metric of Eqs. (\ref{8a}) and 
(\ref{8b}) is not only not at all flat, it even possesses a Riemann tensor which 
is explicitly non-zero (and thus explicitly not flat) even in this very same 
lowest order in $g$ (or $\gamma$) under which Eq. (\ref{9}) was actually 
derived. The answer to this puzzle is that while the Christoffel symbols are 
first order derivative functions of the metric, the Riemann tensor itself is a 
second order such derivative. Thus to get the lowest non-trivial term in the 
Riemann tensor we need to expand the metric to second order in $x/R,~y/R,~z/R$. 
Since a first order expansion suffices for the Christoffel symbols, we thus see 
that there is actually a  mismatch between orders of expansion of the Christoffel 
symbols and the Riemann tensor. Hence a first order study of the geodesics is 
simply not sensitive to the curvature, and thus we not only see why the 
equivalence principle works for weak gravity near the surface of a system such 
as the earth, we even see why it has to do so.\footnote{With 
$g^{\mu \nu}_{\phantom{\mu\nu};\nu}$ vanishing identically, Riemannian geometries 
possess no non-trivial covariant first order derivative function of the metric at 
all. Consequently, lowest order trajectories can only be described by 
non-tensors such as the coordinate dependent Christoffel symbols, with 
only the sum of the two quantities  $d^2x^{\lambda}/d\tau^2$ and 
$\Gamma^{\lambda}_{\mu \nu}(dx^{\mu}/d\tau) (dx^{\nu }/ d\tau)$ actually 
transforming as a contravariant vector. In lowest order then, the coupling of a 
particle to gravity has to be purely inertial.} Moreover, on recognizing this 
fact, we immediately realize that a typical non-geodesic but still fully
covariant particle equation of motion such as\footnote{This (purely illustrative) 
equation of motion can actually be derived by a stationary variation of the 
action $I=-mc\int d\tau-\kappa_1\int d\tau R^{\beta}_{\phantom {\beta} \beta}$ 
where $\kappa_1$ is an appropriate constant.} 
\begin{eqnarray}
m\left( {d^2x^{\lambda} \over d\tau^2}
+\Gamma^{\lambda}_{\mu \nu}
{dx^{\mu} \over d\tau}{dx^{\nu } \over d\tau} \right)
=-\kappa_1 R^{\beta}_{\phantom {\beta} \beta} 
\left( {d^2x^{\lambda} \over d\tau^2}
+\Gamma^{\lambda}_{\mu \nu}
{dx^{\mu} \over d\tau}{dx^{\nu } \over d\tau} \right)
\nonumber \\
-\kappa_1 R^{\beta}_{\phantom {\beta} \beta ;\alpha}
\left( g^{\lambda \alpha}+                                                      
{dx^{\lambda} \over d\tau}                                                      
{dx^{\alpha} \over d\tau}\right)
\label{15}
\end{eqnarray}                 
will just as readily satisfy the same weak gravity tests as the Eq. (\ref{2}) 
geodesic itself. Moreover, since Eq. (\ref{2}) and Eq. (\ref{15}) both reduce
to the Cartesian coordinate Newtonian law $md^2x^{\lambda}/d\tau^2=0$ in the 
absence of curvature, both equations represent valid curved space 
generalizations of Newton's second law of motion, with Eq. (\ref{15}) possessing
both inertial (the Christoffel symbols) and non-inertial (the Ricci scalar) 
contributions.\footnote{In passing it is perhaps worth stressing that some 
relativists regard the equivalence principle as the statement that gravitational 
effects are strictly inertial, with Eq. (\ref{2}) then being the only possible 
allowable coupling of a particle to gravity. However, we take the far more 
cautious view here that while the Christoffel symbols certainly do possess 
the nice, purely geometric inertial property of being simulatable by an 
acceleration in flat spacetime, that does not, and in fact cannot, preclude there 
being a non-inertial, truly coordinate independent, coupling to gravity as well, 
with this issue actually being a dynamical rather than a geometrical one which is 
only decidable by consideration of the curved spacetime field equations with 
which real (as opposed to test) particles are associated.} With Eqs. (\ref{2}) 
and (\ref{15}) having very different strong gravity continuations, we 
thus see that the weak gravity successes of the equivalence principle reveal 
nothing about how Eq. (\ref{2}) would in fact fare in strong gravity, with an 
Eotvos experiment near the surface of a black hole not being at all guaranteed to 
give a null result. Since the presence of such non-inertial terms is actually 
expected\cite{Mannheim2000} to be the general rule rather than the exception in 
field theory,\footnote{The electromagnetic vector potential equation of motion 
$g^{\alpha \beta} A_{\mu;\alpha; \beta}-A^{\alpha}_{\phantom{\alpha};\alpha ;\mu}
+A^{\alpha}R_{\mu \alpha}=0$ contains an explicit non-inertial piece in 
curved spacetime, with a similar situation being found for both the Dirac field 
(vierbeins being non-inertial)  and for the non-minimally coupled scalar field 
$S(x)$ with equation of motion $S^{;\mu}_{\phantom{;\mu};\mu}-(mc/\hbar)^2S 
+\xi R^{\alpha}_{\phantom {\alpha} \alpha}S/6=0$ where $\xi$ is a dimensionless 
parameter.} it would appear that particle motions may not in fact be controlled 
by the strong gravity event horizons and trapped surfaces associated with 
Schwarzschild metric geodesic motion after all, with gravity itself possibly 
being able to protect particles from such phenomena in the strong gravity 
limit,\footnote{Not only could non-inertial effects such as those exhibited in 
Eq. (\ref{15}) become significant in strong gravitational fields, there could 
even be a switch over to the effective repulsion associated with 
$B(r)=1+\gamma r$ type metrics as particles are accelerated to high velocities.} 
with the strong gravity extrapolation of standard weak gravity thus being another 
potentially unreliable extrapolation.  

\section{The Case for Repulsive Gravity}

Recently, through study of type 1A supernovae at very high redshift 
\cite{Riess1998,Perlmutter1998}, it has become possible to explore the attractive 
or repulsive character of gravity on cosmological distance scales, with it now 
being possible to determine whether the universe itself might indeed be slowing 
down or whether it might perhaps actually be speeding up. In terms of the 
standard Robertson-Walker cosmological line element
\begin{equation}
d\tau^2 =c^2dt^2-R^2(t)[(1-kr^2)^{-1}dr^2+r^2d\Omega]
\label{16}
\end{equation}
with associated scale factor $R(t)$ and spatial curvature $k$, the new high $z$ 
data have made it possible to extend Hubble plot measurements of the 
temporal behavior of $R(t)$ beyond lowest order in time, with the current era 
value of the $q(t_0)=-\ddot{R}(t_0)R(t_0)/\dot{R}^2(t_0)$ deceleration parameter 
having now become as amenable to observation as the current value of the  
$H(t_0)=\dot{R}(t_0)/R(t_0)$ Hubble parameter itself. Moreover, the actual 
observations themselves have now provided the first direct evidence that gravity 
might actually contain an explicit repulsive component. Specifically, in terms of 
the standard Einstein-Friedmann cosmological evolution equations, viz. 
\begin{equation}
\dot{R}^2(t) +kc^2=\dot{R}^2(t)(\Omega_{M}(t)+\Omega_{\Lambda}(t))
\label{17}
\end{equation}
and 
\begin{equation}
q(t)=(n/2-1)\Omega_{M}(t)-\Omega_{\Lambda}(t)=
(n/2-1)(1+kc^2/\dot{R}^2(t)) - n\Omega_{\Lambda}(t)/2 
\label{18}
\end{equation}
where $\Omega_{M}(t)=8\pi G\rho_{M}(t)/3c^2H^2(t)$ is due to ordinary matter 
(viz. matter for which $\rho_{M}(t)=A/R^n(t)$ where $A>0$ and $3\leq n \leq 4$) 
and where $\Omega_{\Lambda}(t)=8\pi G\Lambda/3cH^2(t)$ is due to a possible 
cosmological constant $\Lambda$, it was found \cite{Riess1998,Perlmutter1998} 
that the data constrained the allowable current (n=3) era values of the 
parameters $\Omega_{M}(t_0)$ and $\Omega_{\Lambda}(t_0)$ to a quite small region 
in which $\Omega_{\Lambda}(t_0)\simeq\Omega_{M}(t_0)+1/2$ or so, with (the 
presumed positive) $\Omega_{M}(t_0)$ being limited to the range $(0,1)$ or so and 
with $\Omega_{\Lambda}(t_0)$ being limited to the range $(1/2,3/2)$ or so, to 
thus yield a current era deceleration parameter $q(t_0)$ which had to lie within 
the expressly negative $(-1/2,-1)$ interval. While these data thus appear to 
point toward a universe which is actually currently accelerating, systematic 
effects (such as an apparent evolutionary effect between high and low $z$ 
supernovae \cite{Riess2000}) are actually large enough that the data could still 
support a positive value for $q(t_0)$, albeit one which would however still have 
to be substantially smaller than the $q(t_0)=1/2$ value expected in the standard 
$\Omega_{k}(t_0)=-kc^2/\dot{R}^2(t_0) \simeq 0$ flat inflationary universe 
paradigm \cite{Guth1981} in which $\Omega_{M}(t_0)=1$ and $\Omega_{\Lambda}(t_0)
=0$. Thus even if the universe is not accelerating, there would still have to be 
some cosmic repulsion with respect to standard inflation (i.e. with respect to 
the expressly positive $q(t_0)=(n/2-1)\Omega_{M}(t_0)\equiv 1/2$ associated with 
normal, gravitationally attractive matter in a flat universe), even if this 
needed cosmic repulsion did not actually dominate over the positive 
$\Omega_{M}(t_0)$ contribution and lead to a net overall acceleration. The data 
thus entail the existence of some form or other of repulsive component to 
gravity on cosmological distance scales, and it is to the possibility that 
gravity need not always be strictly attractive which we therefore now turn. 

From a purely phenomenological viewpoint, Eq. (\ref{18}) immediately suggests two 
fairly straightforward ways in which $q(t_0)$ could in fact be reduced below one 
half in standard gravity, viz. a positive $\Omega_{\Lambda}(t_0)$ or a negative 
spatial curvature $k$. Of these proposals the possibility of a non-vanishing
$\Omega_{\Lambda}(t_0)$ was first raised by Einstein himself as long ago as the 
very early days of relativity. Specifically, motivated by the desire to have 
a static universe, he noted that if he modified the Einstein-Hilbert action 
by the addition of a fundamental cosmological constant term, the associated 
cosmology would then admit of a static solution (with $\dot{R}(t)=0$, 
$\ddot{R}(t)=0$ and fixed $R(t)=R_0$ in the Robertson-Walker language) provided
$\rho=A/R_0^3=2c\Lambda=kc^4/4\pi G R_0^2$, i.e. provided $\Lambda$ and $k$ were 
both taken to be positive. In such a solution a positive cosmological constant 
term would act repulsively to counteract the attraction (i.e. deceleration) 
associated with positive $\rho$ and positive $k$, to thus yield a (closed) static 
universe. While such a cosmology quickly fell into disfavor following the 
discovery of the cosmic recession of the nebulae shortly thereafter, it did raise for 
the first time the issue of cosmic repulsion, while also raising a problem that 
has been with us ever since, the notorious cosmological constant problem, a 
problem whose modern variant is not so much one of the possible presence of a 
fundamental macroscopic $\Lambda$ but rather of the possible presence of a 
(potentially unacceptably large) microscopically induced one instead.

As regards the second way to get cosmic repulsion, viz. negative curvature, we 
note that such a mechanism is not in fact tied to any specific cosmological 
evolution equation such as the Einstein-Friedmann one of Eqs. (\ref{17}) and 
(\ref{18}), with it actually turning out to be quite general in nature. 
Specifically, it was shown \cite{Mannheim1998} that, no matter what the explicit 
form of the gravitational field equations of motion themselves, the propagation 
of waves such as Maxwell waves in a general curved spacetime Robertson-Walker 
background is completely analogous to the propagation of flat spacetime Maxwell 
waves in a material medium, with vector (and also scalar) curved space waves 
being found to have a dispersion relation of the form 
$\omega^2/c^2=\lambda^2+k$ in a Robertson-Walker background. Thus we see 
that spatial curvature acts just like a frequency dependent refractive index of 
the form $n(\omega)=c\lambda/\omega=(1-kc^2/\omega^2)^{1/2}$, with the group 
velocity associated with energy transport then being given by $v_g=d\omega/d
\lambda=c(1-kc^2/\omega^2)^{1/2}$. When $k$ is negative, spatial curvature is 
then seen to act as a tachyonic mass, to thus effectively give faster than light 
propagation, with a negative curvature space then acting just like a diverging 
dispersive medium wherein group velocities are explicitly greater than their 
values in empty space. Particles propagating in a negative curvature space are 
thus effectively accelerated (cf. a diverging lens), while those traveling in a 
positive curvature space are accordingly decelerated (cf. a converging lens). 
With the standard wisdom regarding the aftermath of the big bang being that the 
mutual attraction of the galaxies would serve to slow down the expansion of the 
universe (i.e. $q(t)=\Omega_{M}(t)/2>0$), we now see that this particular
wisdom only applies if the galaxies are propagating in an inert, empty space 
(viz. a $k=0$ plane lens) which itself has no dynamical consequences. However, 
once $k$ is non-zero, the galaxies would instead then propagate in a non-trivial 
geometric medium, a medium which can then explicitly participate dynamically, 
with the gravitational field itself which is then present in the medium being 
able to accelerate or decelerate the galaxies according to the sign of its 
spatial curvature. Thus we again see that the standard notion of purely 
attractive gravity needs to be reconsidered once the cosmological curvature $k$ 
is non-zero, with gravity itself (viz. curvature) potentially being able to 
generate some repulsion all on its own.

While both of the above cosmic repulsion mechanisms can readily be utilized 
in standard gravity in order to provide a purely phenomenological fit to the 
supernovae data, nonetheless the specific values explicitly required of the 
parameters $\Omega_{M}(t_0)$ and $\Omega_{\Lambda}(t_0)$ actually pose a 
somewhat severe theoretical problem for the standard theory. Specifically, 
given the radically differing temporal behaviors of $\Omega_{\Lambda}(t)$ and 
$\Omega_{M}(t)$, the apparent current closeness to one of their ratio 
\cite{Riess1998,Perlmutter1998} entails that in the early 
universe this same $\Omega_{\Lambda}(t)/\Omega_{M}(t)$ ratio would have had to 
have been fantastically small (typically of order as small as $10^{-60}$), with 
a Friedmann universe only being able to evolve into the currently observed one 
if this ratio had been extremely fine-tuned by fixing the initial conditions in 
the early universe to incredible accuracy. Moreover, even if we take advantage 
of the systematic uncertainties identified in \cite{Riess2000} which might  
potentially (but not necessarily) permit us to set the current value of 
$\Omega_{\Lambda}(t_0)$ to zero, since $\Omega_{M}(t_0)$ would still be required 
to be less than one even in such a case, Eq. (\ref{17}) would then 
oblige the current value of $\Omega_{k}(t_0)$ to be necessarily different from 
zero. Then, given the different temporal behaviors of $\Omega_{M}(t)$ and 
$\Omega_{k}(t)$, this time it would be the $\Omega_{k}(t)/\Omega_{M}(t)$ ratio 
which would need early universe fine-tuning. With Eq. (\ref{17}) being writable 
as $\Omega_{M}(t)+\Omega_{\Lambda}(t)+\Omega_{k}(t)=1$, we thus see that current 
era non-vanishing of any two of these three quantities entails some 
form of early universe fine-tuning problem. Since the current data do not 
support the one point (viz. the $\Omega_{M}(t_0)=1$, $\Omega_{\Lambda}(t_0)=0$, 
$\Omega_{k}(t_0)=0$ inflationary universe) which could be reached without a 
fine-tuning of Eq. (\ref{17}), we see that, even with the current observational 
uncertainties, the new high $z$ data will not support a Friedmann cosmology 
without some form or other of fine-tuning problem, with the non-vanishing of 
$1-\Omega_{M}(t)$ entailing the explicit presence of some cosmic repulsion. 

Moreover, apart from the above macroscopic fine-tuning problems, microscopic 
quantum physics presents cosmology with yet more problems, with the current value 
of the $\Omega_{\Lambda}(t_0)/\Omega_{M}(t_0)$ ratio actually being expected to 
be absolutely enormous - potentially of order $10^{120}$ if generated by quantum 
gravity, and of typical order $10^{60}$ if generated by elementary particle 
physics phase transitions such as the electroweak one.\footnote{Associating a 
typical temperature scale $T_V$ with the electroweak vacuum breaking phase 
transition and a temperature $T(t)$ with the ordinary matter in the universe 
leads to $\Omega_{\Lambda}(t)/\Omega_{M}(t)=T_V^4/T^4(t)$, a ratio which is 
of order $10^{60}$ today.} Microscopic physics thus leads to an expectation 
which is nowhere near the current data at all, with this then being the modern 
variant of the cosmological constant problem to which we referred 
earlier.\footnote{Moreover, should it turn out that there is no 
phenomenological need for a non-zero $\Omega_{\Lambda}(t_0)$ after all, even 
then we would still have to explain why $\Omega_{\Lambda}(t_0)$ is not as big 
as its theoretical expectation, 
with the disappearance of the need not entailing the disappearance of the 
problem.} In all then we thus identify an uncomfortably large number of problems 
for current cosmology (even if explicitly less than one, an $\Omega_{M}(t_0)$ of 
order one would still require an enormous amount of as yet totally undetected 
non-luminous, expressly non-baryonic, cosmological dark matter), and see that 
they all appear to have one common ingredient, namely the use of the evolution 
equation of Eq. (\ref{17}) in the first place. Thus it is highly suggestive 
\cite{Mannheim1998,Mannheim1999} that the problems that the standard cosmology 
currently faces might all derive from the lack of reliability of the 
extrapolation of standard gravity wisdom beyond its solar system origins. In the 
following then we shall thus relax this assumption, and in particular we shall 
show that all of the above problems can readily be resolved if gravity acquires 
one further form of cosmic repulsion, namely that due to an effective 
cosmological Newton constant which is expressly taken to be 
negative.

In trying to identify the root cause of the above problems we note that the 
macroscopic Friedmann equation fine-tuning problem arises because of mismatch 
between the early and current universes, while the microscopic cosmological 
constant problem arises because of a clash between elementary particle physics 
and gravitational physics, with particle physics wanting a large $\Lambda$ and 
gravity a small one. Since this clash is between different branches of physics,
we should not immediately assume that it is the particle physics which is at 
fault. Rather, the indications of particle physics might well be correct, with 
its contribution to $\Lambda$ actually being as big as it would appear to be. 
Indeed, the very failure to date of attempts to quench the particle physics 
$\Lambda$ from so large an expected value might even be an indicator that it is 
not in fact quenched, with it being reasonable to then ask what the implications 
for cosmology are if $\Lambda$ really is big, and whether cosmology could 
actually accommodate a large $\Lambda$. To this end we note that the quantity 
which is of relevance to cosmological evolution is not in fact $\Lambda$ itself 
but rather $\Omega_{\Lambda}(t)=8\pi G\Lambda/3cH^2(t)$, i.e. not the energy of 
the vacuum itself, but rather, its contribution to gravitational evolution, with 
only this latter quantity being observable. With this latter contribution 
depending not just on $\Lambda$ but also on $G$, we see that a quenching of 
$\Omega_{\Lambda}(t)$ could potentially be achieved not by quenching $\Lambda$ 
but by quenching $G$ instead. Thus again we are led to consider $G$ as being 
only an effective parameter, one whose cosmological coupling might be altogether 
smaller than that relevant to the solar system. We shall thus explore the 
possibility that the effective cosmological $G$ is both small and negative, 
first as a general effect and then in an explicit solvable model.

As regards explicitly trying to find a solution to the Friedmann universe 
fine-tuning problem, we note that since the standard cosmology has a big bang, 
the early universe $\dot{R}(t)$ would have to be divergent at $t=0$ (or at least 
be very large), with Eq. (\ref{17}) then requiring the quantity $\Omega_{M}(t=0)
+\Omega_{\Lambda}(t=0)$ to be equal to one at $t=0$, regardless of what 
particular value the spatial curvature $k$ might take. Then, given the radically 
different temporal behaviors of $\Omega_{M}(t)$, $\Omega_{\Lambda}(t)$ and 
$\Omega_{k}(t)$, we see directly that no cosmology which obeys this initial 
constraint, be it flat or non-flat, could ever evolve into one in which 
$\Omega_{\Lambda}(t_0) \simeq \Omega_{M}(t_0) \simeq O(1)$ today (or into one 
in which $\Omega_{M}(t_0) \simeq O(1)$ should $\Omega_{\Lambda}(t)$ just happen 
to be zero) without extreme fine tuning. Hence it is the big bang itself which 
is creating the fine-tuning problem, with it being very difficult for a 
Friedmann universe to evolve from a singular early state into the one currently 
observed. In order to eliminate such an incompatibility  we are thus led to 
consider removing the big bang singularity from cosmology altogether, and have 
the universe expand from some initial (but still very hot) state characterized 
by $\dot{R}(t=0)=0$ instead.\footnote{Once $\dot{R}(t=0)$ is non-singular, the 
natural definition of the initial time in a universe which expands is then the 
one where $R(t)$ is at a minimum, with its minimum value not needing to be zero 
in a non-singular cosmology.} Since the big bang singularity itself derives from 
the assumption that gravity continues to be attractive even when it is strong, 
we are thus led to consider the possibility that the effective cosmological $G$ 
actually be repulsive, with cosmology then being able to protect itself from 
its own singularities and potentially rid itself of fine-tuning 
problems.\footnote{That the origin of the flatness problem could be traced to 
the positivity of $G$ was already noted quite some time ago in 
\cite{Mannheim1992}, where it was also pointed out that a negative effective $G$ 
repulsive cosmology would actually have no fine-tuning flatness problem at all, 
with quantities which would have had to cancel to very high accuracy in Eq. 
(\ref{17}) no longer having to do so.} 

Continuing in this same vein, we note further that if the universe turns out to 
ultimately be an accelerating one (either even already, or at some time later in 
the future\footnote{Even if $\Omega_{\Lambda}(t_0)$ turns out to be negligible 
today, as long as it is greater than zero, then no matter by how little, there
will eventually come a time when $\Omega_{\Lambda}(t)$ will ultimately come 
to dominate the expansion of the universe.}), $\dot{R}(t)$ will then become 
arbitrarily large in the late (rather than the early) universe, with the 
quantity $\Omega_{M}(t)+\Omega_{\Lambda}(t)$ tending to one at late times, 
again independent of the value of $k$. Then, in such a late universe it would be 
$\Omega_{\Lambda}(t)$ which would have to tend to one, no matter what may or may 
not have happened in the early universe, and regardless of how big $\Lambda$ 
itself might actually be. Thus at very late times the cosmological constant 
problem would not only get solved, it would get solved by cosmology itself, with 
an accelerating universe always being able to quench $\Omega_{\Lambda}(t)$ once 
given time enough to do so. Thus we see that the very same problem which is 
generated by the very existence of a cosmological constant is then made solvable 
by the very cosmic acceleration which it simultaneously produces; with the key 
question for cosmology thus being whether the universe is already sufficiently 
late for this quenching to have already taken place. However, in order for this 
to actually be the case, it is necessary that the current era contribution of 
ordinary matter to the evolution of the universe be cosmologically insignificant 
(i.e. that $\Omega_{M}(t)=8\pi G\rho_{M}(t)/3c^2H^2(t)$ already be close to an 
asymptotically expected value of zero). Thus again we find ourselves led to 
considering the possibility that the effective cosmological $G$ be altogether 
smaller than the one relevant in the solar system, so that, no matter how big 
$\rho_{M}(t)$ itself might be, ordinary matter would then no longer be of 
relevance to the current expansion, though, given its temporal behavior, 
$\rho_{M}(t)$ would still be highly significant at altogether earlier 
times.\footnote{Thus we do not seek to change the matter content of the 
universe, but rather only to modify how it impacts on cosmological evolution. 
Moreover, with the ratio $\Omega_{\Lambda}(t)/\Omega_{M}(t)=T_V^4/T^4(t)$ 
actually being independent of $G$, we see that the requirement that 
$T_V/T(t_0)$ be large and that the current era $\Omega_{\Lambda}(t_0)$ be of 
order one together entail that $\Omega_{M}(t_0)$ be highly suppressed, something 
readily achievable if the effective cosmological $G$ is then very small.} 
Thus to conclude, we see that many outstanding cosmological puzzles are readily 
addressable if gravity were indeed to be describable cosmologically by a 
repulsive and very small effective $G$, and thus motivated we shall, in the 
following, present a specific alternate gravitational theory, actually the 
conformal gravity theory to which we referred above, where this is precisely 
found to be the case.

\section{Repulsive Gravity}  
   
Since we are not currently aware of how it might be possible to implement the 
above general ideas within standard gravity (a $G$ which evolves with 
temperature from the early universe until today is certainly conceivable within 
standard gravity though perhaps not one which might also change sign as it 
evolves), we shall instead turn to an alternate gravity theory, viz. conformal 
invariant gravity, a theory which is immediately suggested since gravity then 
becomes a theory with dimensionless coupling constants and no intrinsic
mass  scales just like the three other fundamental interactions. Thus gravity 
becomes a theory which is power counting renormalizable, with the absence of any 
intrinsic mass scale immediately obliging any fundamental cosmological constant 
term to be set to zero in it,\footnote{The absence of any intrinsic quantum 
gravity scale not merely eliminates the Planck length from quantum relevance, it 
also provides for a theory of gravity in which quantum gravity fluctuations 
themselves cannot then generate a cosmological constant term at all; with any 
induced microscopic $\Lambda$ then only being generatable by elementary particle 
physics phase transitions.} to thus yield \cite{Mannheim1990} a theory in which 
the cosmological constant term is actually controlled  by an underlying symmetry, 
an objective long sought in the standard model. With the conformal theory also, 
as noted earlier, possessing a good Newtonian limit despite the absence of any 
intrinsic Einstein-Hilbert term (the conformal gravitational action is uniquely 
given by the conformal $I_W$ action of Eq. (\ref{3})), we see that a possible 
reason why the standard gravity cosmological constant problem has resisted 
solution for so long could be that was always being sought was a symmetry which 
would eliminate a fundamental $\Lambda$ but not $I_{EH}$ rather than one 
which would in fact eliminate both. Since solar system information had never in 
fact required the presence of $I_{EH}$ in the first place, we see that its 
removal actually opens up a new line of attack on the cosmological constant 
problem; and as we shall now see, the conformal symmetry not merely controls any 
fundamental cosmological constant term, even after the conformal symmetry is 
spontaneously broken (something needed to generate particle masses in the 
conformal symmetry), the continuing tracelessness of the energy-momentum tensor 
sharply constrains the magnitude of any induced one in a manner which is then 
found to automatically implement the $\Omega_{\Lambda}(t)$ quenching mechanism we 
introduced above. 

The cosmology associated with conformal gravity was first presented in 
\cite{Mannheim1992} where it was shown, well in advance of the recent 
discovery of cosmic repulsion, to be one with an effective repulsive 
cosmological $G$ just as desired above. To discuss conformal cosmology it is 
convenient to consider the conformal matter action
\begin{equation} 
I_M=-\hbar\int d^4x(-g)^{1/2}[S^\mu S_\mu/2
- S^2R^\mu_{\phantom{\mu}\mu}/12+\lambda S^4
+i\bar{\psi}\gamma^{\mu}(x)(\partial_\mu+\Gamma_\mu(x))\psi
-gS\bar{\psi}\psi] 
\label{19}
\end{equation}
where we take the elementary particle physics matter fields to be generically
represented by massless fermions for definitiveness and simplicity, with a
conformally coupled massless scalar field being introduced to serve as the
order parameter associated with spontaneous breakdown of the scale symmetry.
With dynamical symmetry breaking being needed in order to generate elementary
particle masses in the scaleless interacting massless fermion and gauge 
boson theories now standard in particle physics, the scalar field
$S(x)$ introduced here should be thought of not as a fundamental scalar field
but as the expectation value of an appropriate fermion multilinear in an
appropriate coherent fermionic state. $S(x)$ is thus to serve as a cosmological
analog of the Cooper pair of superconductivity theory, with the action of Eq.
(\ref{19}) serving as an analog of the Ginzburg-Landau phenomenological
superconductivity Lagrangian. In such a case the vacuum energy would be zero
above the critical temperature where the order parameter $S(x)$ would then
vanish, and would be expressly negative below it. Simulating the vacuum
energy below the critical point by the effective $\hbar\lambda S^4(x)\equiv 
\Lambda$  term, then entails (as noted in \cite{Mannheim1999}) that in a theory 
which is scale invariant above the critical point, the effective
$\Lambda$ which is induced below the critical point is then expressly
negative. Thus unlike the situation in the standard theory, in the conformal
case the sign of the induced cosmological constant term is explicitly determined. 

For the above matter action the matter field equations of motion take the form  
\begin{eqnarray}
 i\gamma^{\mu}(x)[\partial_{\mu} +\Gamma_\mu(x)]\psi 
- g S \psi =0 
\nonumber \\
S^\mu _{\phantom{\mu};\mu}
+ SR^\mu_{\phantom{\mu}\mu}/6-
4\lambda S^3+ g\bar{\psi}\psi=0
\label{20}
\end{eqnarray}
with the matter energy-momentum tensor being given by 
\begin{eqnarray}
T^{\mu \nu}=\hbar \{i \bar{\psi} \gamma^{\mu}(x)[ 
\partial^{\nu}                    
+\Gamma^\nu(x)]\psi
+2S^\mu S^\nu/3 
-g^{\mu\nu}S^\alpha S_\alpha/6 
- SS^{\mu;\nu}/3
\nonumber \\             
+ g^{\mu\nu}SS^\alpha_{\phantom{\alpha};\alpha}/3                              
- S^2(R^{\mu\nu}-
g^{\mu\nu}R^\alpha_{\phantom{\alpha}\alpha}/2)/6
-g^{\mu \nu}\lambda S^4 \}.  
\label{21}
\end{eqnarray}
Thus, when the scalar field acquires a non-zero vacuum expectation value $S_0$, 
the energy-momentum tensor then takes the form (for a perfect matter fluid 
$T^{\mu\nu}_{kin}$ of the fermions) 
\begin{equation}
T^{\mu\nu}=T^{\mu\nu}_{kin}-\hbar S_0^2(R^{\mu\nu}-
g^{\mu\nu}R^\alpha_{\phantom{\alpha}\alpha}/2)/6            
-g^{\mu\nu}\hbar\lambda S_0^4.
\label{22}
\end{equation}
Since the Weyl tensor $C^{\lambda \mu \nu \kappa}$ and the quantity 
$W^{\mu \nu}$ of Eq. (\ref{5}) both vanish identically in the highly symmetric 
Robertson-Walker geometry, the complete cosmological solution to the joint 
scalar, fermionic, and gravitational field equations of motion then reduces 
\cite{Mannheim1998} to just one relevant equation, viz. 
\begin{equation}
T^{\mu \nu}=0, 
\label{23}
\end{equation}
a remarkably simple condition which immediately fixes the zero of energy. Thus 
in its spontaneously broken phase conformal cosmology is described by the 
equation
\begin{equation}
\hbar S_0^2(R^{\mu\nu}-
g^{\mu\nu}R^\alpha_{\phantom{\alpha}\alpha}/2)/6 =T^{\mu\nu}_{kin}           
-g^{\mu\nu}\hbar\lambda S_0^4,
\label{24}
\end{equation}
an equation which we recognize as being none other than that of standard gravity
save only that the quantity $-\hbar S_0^2 /12$ has replaced the familiar 
$c^3/16 \pi G$. Conformal cosmology thus acts exactly like a standard gravity
theory in which $G$ is effectively negative, with its magnitude actually becoming 
smaller the larger $S_0$ gets to be, i.e. the same particle physics mechanism 
which makes the cosmological constant large serves to make the effective 
cosmological $G$ small. Thus, just as desired, conformal cosmology is controlled 
by an effective $G$ which is both negative and small. Noting further that the 
Weyl tensor vanishes in high symmetry, cosmologically relevant geometries such 
as the homogeneous Robertson-Walker one, but not in low symmetry ones such as 
Schwarzschild (viz. geometries which are generated by the presence of local 
spatial inhomogeneities in an otherwise homogeneous cosmological background),
we  thus see that the gravitational coupling constant $\alpha_g$, viz. the
one which  according to Eq. (\ref{8}) explicitly controls the geometry
outside of a local static source, simply decouples from cosmology. Thus in
conformal gravity  (inhomogeneous) locally attractive and (homogeneous)
globally repulsive gravity  can readily coexist, with solar system physics
indeed not then being a good  guide to the behavior of gravity in altogether
different circumstances. 

Given Eq. (\ref{24}) the conformal cosmology evolution equations immediately
take the form 
\begin{equation}
\dot{R}^2(t) +kc^2 =
-3\dot{R}^2(t)(\Omega_{M}(t)+
\Omega_{\Lambda}(t))/ 4 \pi S_0^2 L_{PL}^2
\equiv \dot{R}^2(t)(\bar{\Omega}_{M}(t)+
\bar{\Omega}_{\Lambda}(t))
\label{25}
\end{equation}
and 
\begin{equation}
q(t)=(n/2-1)\bar{\Omega}_{M}(t)-\bar{\Omega}_{\Lambda}(t)
\label{26}
\end{equation}
(Eq. (\ref{25}) serves to define $\bar{\Omega}_{M}(t)$ and 
$\bar{\Omega}_{\Lambda}(t)$), and are thus remarkably similar to the standard
evolution equations of Eqs. (\ref{17}) and (\ref{18}); with our entire earlier 
general discussion on the need to replace the cosmological $G$ by a small 
negative effective one now finding explicit realization in a specific alternate 
gravitational theory. With both the effective $G$ and the vacuum breaking 
$\Lambda$ being negative in conformal gravity, we thus see 
that $\bar{\Omega}_{M}(t)$ is necessarily negative while 
$\bar{\Omega}_{\Lambda}(t)$ is positive. Consequently, the ordinary matter energy 
density and the vacuum energy density both naturally lead to cosmological 
repulsion, with the conformal $q(t)$ always having to be less than or equal to 
zero. Conformal cosmologies thus never decelerate with each epoch seeing some 
measure of cosmic repulsion no matter what the explicit magnitudes of 
$\bar{\Omega}_{M}(t)$ and $\bar{\Omega}_{\Lambda}(t)$ might be.

In order to determine how much repulsion there might be in any given epoch
it is necessary to determine the value of the spatial curvature $k$. To this
end we note while we cannot immediately fix $k$ from study of the broken
symmetry phase itself ($\bar{\Omega}_{M}(t)$ and $\bar{\Omega}_{\Lambda}(t)$ 
make contributions of opposite sign in Eq. (\ref{25})), it is possible to 
extract information about $k$ from the high temperature phase above all phase 
transitions, a phase where the order parameter vanishes, a phase which can be
modeled entirely by  the presence of just a perfect fluid, viz. one in which
the entire $T^{\mu\nu}$  is given by $T^{\mu \nu}_{kin}$. In such a high
temperature Robertson-Walker phase the gravitational equations of motion of Eq.
(\ref{4}) reduce to the condition
\begin{equation}
T^{\mu \nu}_{kin}=0, 
\label{27}
\end{equation}
an equation which quite remarkably actually has a non-trivial solution in curved 
space. 

To explore such a solution, we note that while we have generically identified 
the fields in $T^{\mu\nu}_{kin}$ to be perfect fluid fermions, for calculational 
purposes it is more convenient to consider them to be non-interacting scalar 
fields instead.  We thus populate the non-spontaneously broken high temperature 
universe by a perfect fluid built out of the modes of a normal 
(i.e. not spontaneously broken) scalar field with equation of motion              
\begin{equation}   
S^\mu _{\phantom{\mu};\mu}
+ SR^\mu_{\phantom{\mu}\mu}/6=0
\label{28}
\end{equation}
and energy-momentum tensor
\begin{equation}   
T^{\mu \nu}=2S^\mu S^\nu/3 -g^{\mu\nu}S^\alpha S_\alpha/6 
- SS^{\mu;\nu}/3
+g^{\mu\nu}SS^\alpha_{\phantom{\alpha};\alpha}/3                               
- S^2(R^{\mu\nu}- g^{\mu\nu}R^\alpha_{\phantom{\alpha}\alpha}/2)/6.
\label{29}
\end{equation}
With the Ricci scalar being given by the 
spatially independent $R^\alpha_{\phantom{\alpha}\alpha}=
-6(k+R(t)\ddot{R}(t)+\dot{R}^2(t))/R^2(t)$, the scalar field equation of motion
is found to be separable. Thus on introducing the conformal time $p=c\int^t
dt/R(t)$ and on setting $S(x)=f(p)g(r,\theta,\phi)/R(t)$, Eq. (\ref{28}) is
found to reduce to ($\gamma$ denotes the determinant of the spatial metric) 
\begin{equation}   
{1 \over f(p)}\left[ {d^2f \over dp^2} +kf(p) \right]=
{1 \over g(r,\theta,\phi)}\gamma^{-1/2}\partial_i[\gamma^{1/2}\gamma^{ij}
\partial_j g(r,\theta,\phi)]=-\lambda^2,
\label{30}
\end{equation}
where we have introduced a separation constant $-\lambda^2$. With the
dependence on $f(p)$ on $p$ being harmonic, the frequencies thus have to
satisfy $\omega^2=\lambda^2+k$, a relation we actually presented earlier. For 
the spatial dependence we can further set
$g(r,\theta,\phi)=g^{\ell}_{\lambda}(r) Y^{m}_{\ell}(\theta,\phi)$ where
$g^{\ell}_{\lambda}(r)$ obeys the radial equation
\begin{equation}   
\left[ (1-kr^2){\partial^2 \over \partial r^2}+{(2-3kr^2) \over r}
{\partial \over \partial r}-{\ell(\ell +1) \over r^2} +\lambda^2 \right]
g^{\ell}_{\lambda}(r)=0,
\label{31}
\end{equation}
with the radial solutions being given \cite{Mannheim1988} by
$j_{\ell}(\omega r)/\omega$ Bessel functions when $k=0$, by associated Legendre
functions when $k=-1$, viz.
\begin{equation}   
g^{\ell}_{\lambda}(r)=[\pi\omega^2(\omega^2+1^2)...(\omega^2+\ell^2)/2]^{-1/2}
sinh^{\ell}\chi\left[  {d \over d cosh \chi}\right] ^{\ell+1}{cos \omega 
\chi
\over \omega}  
\label{32}
\end{equation}
where $r=sinh \chi$, and by Gegenbauer polynomials when $k=1$, viz.
\begin{equation}   
g^{\ell}_{\lambda}(r)=[\pi\omega^2(\omega^2-1^2)...(\omega^2-\ell^2)/2]^{-1/2}
sin^{\ell}\chi\left[  {d \over d cos \chi}\right] ^{\ell+1}{cos \omega 
\chi
\over \omega}  
\label{33}
\end{equation}
where $r=sin \chi$. With these normalizations (which differ slightly from those
given in \cite{Mannheim1988}), an incoherent averaging of the energy-momentum 
tensor of Eq. (\ref{29}) over all the available spatial states associated 
with a given frequency $\omega$ is then found \cite{Mannheim1988} to lead
directly for every allowed $\omega$ to the traceless kinematic perfect fluid
\begin{equation}   
T^{\mu \nu}_{kin}= {\omega^2(4U^{\mu}U^{\nu}+g^{\mu \nu}) \over 6\pi^2R^4(t)}
\label{34}
\end{equation}
in all three of the spatial geometries.\footnote{On purely general geometric
grounds the most general rank two tensor in a Robertson-Walker geometry is 
writable as $T^{\mu \nu}=(A(t)+B(t))U^{\mu}U^{\nu} +B(t)g^{\mu \nu}$, with the 
coefficients $A(t)$ and $B(t)$ being otherwise unconstrained, so that their ratio 
$w(t)=B(t)/A(t)$ need not be time independent in general. However, for a tensor 
which is both traceless and covariantly conserved, it further follows that 
$3B(t)=A(t)=C/R^4(t)$ where $C$ is a pure constant; with the calculation leading 
to Eq. (\ref{34}) thus being a calculation of the value of this constant. While 
on this point, it is worth noting in passing that it is not automatically the 
case that $w(t)$ is necessarily time independent or that $A(t)$ and $B(t)$ 
necessarily have the same dependence on time in the general Robertson-Walker 
cosmology case. Relations between $A(t)$ and $B(t)$ only follow under specific 
dynamical assumptions. If, for example, $A(t)$ and $B(t)$ contain two or more 
separate components, then even if the separate components are related via 
time independent $w_1=B_1/A_1$, $w_2=B_2/A_2$, it does not necessarily follow 
that $B_1+B_2$ is proportional to $A_1+A_2$. Further, even for the restricted 
case of the kinematic $T^{\mu \nu}_{kin}$ in which $A(t)$ and $B(t)$ are 
both associated with just one single completely standard perfect fluid, it
turns out that even in that case they are still not in fact proportional to
each other at all temperatures. To illustrate this point, consider an ideal
$N$ particle classical gas of particles of mass $m$ in a volume $V$ at a
temperature $T$. For  this system the Helmholtz free energy $A(V,T)$ is given
as exp$[-A(V,T)/NkT]= V\int d^3p$ exp$[-(p^2+m^2)^{1/2}/kT]$, so that the
pressure takes the simple  form $P=-(\partial A/ \partial V)_T=NkT/V$, while
the internal energy $U=A-T(\partial A/ \partial T)_V$ evaluates in terms of 
Bessel functions as $U=3NkT+Nm K_1(m/kT)/K_2(m/kT)$. In the two limits 
$m/kT \rightarrow 0$, $m/kT \rightarrow \infty$ we then find that 
$U \rightarrow 3NkT$, $U \rightarrow Nm+3NkT/2$. Thus only at these two extreme 
temperature limits does it follow that the energy density and the pressure are in 
fact proportional, with their relation in intermediate regimes such as the 
transition region from the radiation to the matter era being far more 
complicated.} With the condition $T^{\mu \nu}_{kin}=0$ of Eq. (\ref{27}) then only 
permitting the soft $\omega=0$ modes, and with only the $k=-1$ radial equation 
admitting of a non-trivial radial solution in such a case (viz. the entire 
infinite set\footnote{In passing we note that it might prove interesting should 
there be a group under which this infinite tower of states transforms 
irreducibly.} of all $\ell$ modes built on the $\ell=0$ solution 
$g^0_1(r)\simeq\chi/sinh \chi$), 
we see that it is possible to satisfy the condition $T^{\mu \nu}_{kin}=0$ 
non-trivially in the negative spatial curvature conformal cosmology case, with 
the very high temperature universe then being composed of a perfect fluid bath 
of soft modes which non-trivially support a $k<0$ universe. Cosmology thus fixes 
the curvature itself, and does so before the onset of any symmetry breaking 
phase transition at all.\footnote{To show that there actually will be a phase 
transition as the temperature drops requires the development of a detailed 
dynamical model for the coherent correlations that are not present in the 
incoherently averaged perfect fluid limit. However, without knowing the details 
of how long range order actually sets in, nonetheless, use of the perfect
fluid  model at temperatures far above the critical region is sufficient to
fix the  sign of $k$ once and for all.} 

It is also illuminating to derive this result in a slightly different fashion.
Since $T^{\mu \nu}_{kin}$ is a conformal invariant tensor, we could also 
evaluate it by first making the conformal transformation $g_{\mu \nu}(x)
\rightarrow R^{-2}(t)g_{\mu \nu}(x)$ on the Robertson-Walker metric, a 
transformation which brings the geometric line element to the form 
$d\tau^2=dp^2-dr^2/(1-kr^2)-r^2d\Omega=dp^2-\gamma_{ij}dx^idx^j$. For this 
transformed metric the Ricci scalar is given by 
$R^\alpha_{\phantom{\alpha}\alpha}=-6k$, with the incoherently averaged soft 
mode contribution to the energy density then being given by
\begin{equation}
T^{00}_{kin}={1 \over 6} \sum_{\ell,m}
\left[\sum_{i=1}^3\gamma^{ii}|\partial_i (g^{\ell}_1
Y_{\ell}^m(\theta,\phi))|^2 +k|g^{\ell}_1
Y_{\ell}^m(\theta,\phi)|^2\right],
\label{35}
\end{equation}
a quantity which can actually vanish non-trivially provided $k$ is negative.
And indeed, through use of the various completeness relations \cite{Mannheim1988} 
which these modes obey, explicit evaluation of $T^{00}_{kin}$ is then found to 
yield   
\begin{eqnarray}
T^{00}_{kin}={1 \over 6} \sum_{\ell,m}
{(2\ell +1) (\ell -m)! \over 4\pi (\ell+m)!}\left[ 
(1-kr^2)P^m_\ell(cos\theta)^2({dg^{\ell}_1 \over dr})^2
+{1 \over r^2}({d P^m_\ell(cos\theta)\over d \theta})^2(g^{\ell}_1)^2 \right.
\nonumber \\
+\left.{m^2 \over r^2 sin^2\theta} P^m_\ell(cos\theta)^2(g^{\ell}_1)^2
+kP^m_\ell(cos\theta)^2(g^{\ell}_1)^2 \right]
\nonumber \\
={1 \over 24 \pi} \sum_{\ell}(2\ell+1)\left[ (1-kr^2)({dg^{\ell}_1 \over dr})^2
+{\ell(\ell+1) \over r^2}(g^{\ell}_1)^2 +k(g^{\ell}_1)^2 \right]
\nonumber \\
={1 \over 24 \pi} \left[  -{2k \over 3\pi}-{4k \over 3\pi}+{2k \over \pi}
\right]=0.
\label{35a}
\end{eqnarray}
Thus we see that when the spatial curvature is negative, the negative energy 
density then present in the gravitational field completely cancels the positive 
energy density of the matter fields, with gravity itself then being able to fix 
the spatial curvature of the universe.

Having now shown that $k$ is uniquely negative in conformal cosmology (we
will show below how galactic rotation curve data actually enable us
to measure $k$ to find that it is indeed negative), we can now proceed to
study its cosmological implications as the universe cools. Thus, with the signs 
of $k$ and $\lambda$ now being fixed (we simulate the negativity of 
$\Lambda=\hbar\lambda S^4_0$ by a negative $\lambda$), Eq. (\ref{25}) is then 
found (in the simpler to treat high temperature era where 
$cT^{00}_{kin}=\rho_{M}(t)=A/R^4=\sigma T^4$) to admit of the unique solution
\cite{Mannheim1998,Mannheim1999} 
\begin{equation}
R^2(t,\alpha>0,k<0)= -k(\beta-1)/2\alpha
-k\beta sinh^2 (\alpha^{1/2} ct)/\alpha
\label{36}
\end{equation}
where have introduced the positive parameter $\alpha =-2\lambda S_0^2$ and the 
parameter $\beta =(1- 16A\lambda/k^2\hbar c)^{1/2}$ which is greater than one. 
In this solution the deceleration parameter is found to take the requisite 
non-positive form
\begin{equation}
q(t,\alpha>0,k<0)=
-tanh^2(\alpha^{1/2}ct)
+2(1-\beta)cosh(2\alpha^{1/2}ct)/
\beta sinh^2(2\alpha^{1/2}ct).
\label{37}
\end{equation}
We thus see that the cosmology is non-singular, having a finite minimum radius 
and an initial $\dot{R}(t=0)$ which is zero rather than infinite. Since the 
cosmology has a minimum radius, it also has a finite maximum temperature 
$T_{max}$ in terms of which Eq. (\ref{36}) can be rewritten as
\begin{equation}
T_{max}^2(\alpha > 0,k<0)/T^2(t,\alpha>0,k<0)=
1+2\beta sinh^2 (\alpha^{1/2} ct)/(\beta-1),
\label{38}
\end{equation}
with the Hubble parameter then being given as 
\begin{equation}
H(t)=\alpha^{1/2}c(1-T^2(t)/T^2_{max})/tanh(\alpha^{1/2}ct).
\label{40b}
\end{equation}
In terms of the convenient effective temperature $T_V$ defined via 
$-c\Lambda=-c\hbar\lambda S^4_0=\sigma T_V^4$, we find that the 
parameter $\beta$ can be expressed as   
\begin{equation}
\beta= (1+T_V^4/T_{max}^4)/(1-T_V^4/T_{max}^4)
\label{39}
\end{equation}
(with $T_V$ thus being less than the maximum temperature $T_{max}$), with the 
temporal evolution of the theory then being given by
\begin{equation}
T_{max}^2/T^2(t)=1+(1+T_{max}^4/T_V^4)sinh^2 (\alpha^{1/2} ct)
\label{40a}
\end{equation}
and with the energy density terms then being given by
\begin{eqnarray}
\bar{\Omega}_{\Lambda}(t)=(1-T^2(t)/T_{max}^2)^{-1}(1
+T^2(t)T_{max}^2/T_V^4)^{-1},
\nonumber \\
\bar{\Omega}_M(t)=-(T^4(t)/T_V^4)\bar{\Omega}_{\Lambda}(t),
\nonumber \\
\Omega _k(t)=-kc^2/\dot{R}^2(t)=1-\bar{\Omega}_M(t)-\bar{\Omega}_{\Lambda}(t).
\label{41}
\end{eqnarray}
With $(1+T^2(t)T_{max}^2/T_V^4)^{-1}$ being a quantity which is always 
necessarily bounded between zero and one (no matter what the magnitude of $T_V$), 
we see \cite{Mannheim1999} that the single, simple requirement that $T_{max}$ be 
very much greater than $T(t_0)$ then entails that $\bar{\Omega}_{\Lambda}(t_0)$ 
must lie somewhere between zero and an upper bound of one today, with its current 
value then being given by $\bar{\Omega}_{\Lambda}(t_0)
=(1+T^2(t_0)T_{max}^2/T_V^4)^{-1}=tanh^2(\alpha^{1/2}ct_0)$ 
according to Eq. (\ref{40a}). Thus, no matter how big $T_V$ might be, 
$\bar{\Omega}_{\Lambda}(t_0)$ must not only be of order one today, it must also 
be approaching its asymptotically expected value of one from below; to thus yield 
a completely natural current era quenching of $\bar{\Omega}_{\Lambda}(t_0)$ 
regardless of the numerical value of any cosmological parameter. Moreover,
the larger $T_V$, the further $\bar{\Omega}_{\Lambda}(t_0)$ must lie below its 
asymptotic bound of one, with it taking a typical value of one-half in the event 
that the current temperature is given by $T(t_0)\simeq T_V^2/T_{max}$, a 
condition which is readily realizable if the conditions $T_{max} \gg T_V$ and 
$T_V \gg T(t_0)$ both hold.\footnote{There are three possible ways in which the 
condition $T_{max} \gg T(t_0)$ can be realized in Eq. (\ref{40a}). If $T_V$ 
is of order $T(t_0)$, $sinh^2 (\alpha^{1/2} ct_0)$ would be very much less than 
one. Similarly, if $T_V$ is of order $T_{max}$, $sinh^2 (\alpha^{1/2} ct_0)$ 
would be very much greater than one. However, if $T_V$ is of intermediate 
order $(T(t_0)T_{max})^{1/2}$, $sinh^2 (\alpha^{1/2} ct_0)$ would then be 
of order one.} Thus we see that $\bar{\Omega}_{\Lambda}(t_0)$ could 
actually be appreciably below one today, and that it would actually fall further 
below one the larger rather than the smaller the cosmological constant, with 
conformal cosmology thus having no difficulty handling a large $T_V/T(t_0)$. Thus 
it is precisely in the event that the cosmological constant is in fact large that 
$\bar{\Omega}_{\Lambda}(t_0)$ is then able to be of a phenomenologically 
acceptable magnitude long before becoming fully asymptotic.\footnote{Since 
$T_{max}$ is greater than $T_V$, a large $T_V/T(t_0)$ entails a large 
$T_{max}/T(t_0)$, and thus a universe old enough for the current era 
$\bar{\Omega}_{\Lambda}(t_0)$ to be given by the bounded 
$tanh^2(\alpha^{1/2}ct_0)$. For large $T_V$ conformal cosmology thus solves the 
cosmological constant problem simply by living a long time.} Beyond this, we note 
additionally, that given the fact that there is an upper bound on 
$\bar{\Omega}_{\Lambda}(t_0)$, we see that a large $T_V/T(t_0)$ will then 
completely quench the current era $\bar{\Omega}_M(t_0)$ altogether. Thus 
in conformal gravity a large cosmological constant naturally leads to current era 
suppression of $\bar{\Omega}_M(t_0)$ just as was desired 
above.\footnote{Phenomenologically, this quenching of $\bar{\Omega}_M(t_0)
=-3\Omega_M(t_0)/4\pi S_0^2 L^2_{Pl}$ requires that the scale factor $S_0$ be 
altogether greater than $L^{-1}_{Pl}$, a condition which is readily realizable 
not only by associating a fundamental temperature with a fundamental $S_0$ which 
is altogether larger than the Planck temperature, but instead, and 
preferentially,  by identifying $S_0$ as a macroscopically occupied order 
parameter \cite{Mannheim1999}, with $S_0$ then being proportional to the (very 
large) number, $N$, of occupied positive energy perfect fluid modes, modes whose 
coherent correlations cause phase transitions to occur in the early universe in 
the first place.} Then, because of this suppression, the current era evolution 
equation of Eq. (\ref{25}) reduces to $1=\Omega_k(t_0)
+\bar{\Omega}_{\Lambda}(t_0)$, with negative curvature explicitly forcing 
$\bar{\Omega}_{\Lambda}(t_0)$ to lie below one. It is thus the negative spatial 
curvature of the universe which bounds and tames the contribution of the 
cosmological constant to cosmology,\footnote{In a $\lambda<0$ cosmology with 
$k>0$, the current era $\bar{\Omega}_M(t_0)$ is still found 
\cite{Mannheim1998,Mannheim1999} to be completely suppressed by large 
$T_V/T(t_0)$, but in such a cosmology $\bar{\Omega}_{\Lambda}(t_0)$ is no longer 
bounded by one from above, but only from below, with the $k>0$ case $T_{max}$ 
being found to be less than $T_V$. Thus we see that in the $k<0$ case it is 
precisely the contribution of negative curvature itself which produces a 
cosmology in which the highest temperature $T_{max}$ is greater than the vacuum 
energy temperature $T_V$ (with there thus being an in principle difference 
between flat space and curved space phase transitions), with curvature itself 
controlling the value of $T_{max}$. Moreover, with there being an explicit 
maximum temperature $T_{max}^2=-k\hbar S_0^2c/2(\sigma A)^{1/2}$ 
in a $k<0$ cosmology even in the absence of any $\lambda S_0^4$ term at all, 
we see that the magnitudes of $T_{max}$ and $T_V$ are fixable independently, 
with the imposition of the condition $T_{max}\gg T_V$ which we use thus being 
readily naturally achievable in the $\lambda \neq 0$ case without the need for 
any fine-tuning of parameters.} with it thus indeed being possible to construct 
a phenomenologically acceptable cosmology in which 
$\Lambda$ can still be as large as particle physics suggests.

From a phenomenological viewpoint, once $\bar{\Omega}_M(t_0)$ is suppressed,
the deceleration parameter is then given by $q(t_0)=-tanh^2(\alpha^{1/2}ct_0)$
(so that it then has to lie between zero and minus one), while the curvature 
contribution is given by $\Omega_k(t_0)=sech^2(\alpha^{1/2}ct_0)$. Thus in 
conformal cosmology first matter dominates the expansion rate (in the early 
universe), then curvature, and finally vacuum energy; and even if the current 
era is not yet vacuum energy dominated, nonetheless $\bar{\Omega}_{\Lambda}(t_0)$ 
will still be under control no matter what the value of $\alpha^{1/2}ct_0$. As 
regards the actual supernovae data themselves, we note that even though the 
phenomenological fitting allowed for solutions with $\Omega_M(t_0)= 0$ 
(typically with $\Omega_{\Lambda}(t_0)=-q(t_0)=1/2$), and even though such 
solutions would not be expected to occur in the standard theory, we see that in 
the conformal theory\footnote{With $\Omega_M(t_0)$ and $\Omega_{\Lambda}(t_0)$ 
being treated as free parameters in supernovae data fitting based on the 
standard Eq. (\ref{17}), those fits are just as equally phenomenological fits 
to the conformal Eq. (\ref{25}).} solutions with $\bar{\Omega}_M(t_0)=0$ and 
$\bar{\Omega}_{\Lambda}(t_0)=1/2$ are right in the region allowed by Eq. 
(\ref{41}). Moreover, in \cite{Riess1998,Perlmutter1998} fits with 
$\Omega_M(t_0)=0$ and $\Omega_{\Lambda}(t_0)=0$ were found, and even though 
those fits were of quality comparable with the best reported 
$\Omega_M(t_0)\neq 0$, $\Omega_{\Lambda}(t_0)\neq 0$ fits, such fits were not 
considered further since in standard gravity they would correspond to an empty 
universe. However, we now see that in conformal gravity not only would such
fits (fits in which $T_V^2/T(t_0)T_{max}\ll 1 $, so that 
$\bar{\Omega}_M(t_0) \simeq0$ $\bar{\Omega}_{\Lambda}(t_0)\simeq 0$) be quite 
acceptable, such fits would not entail an empty universe since it is 
$\bar{\Omega}_M(t_0)=0$ which is suppressed and not $\rho_M(t_0)$ itself. Thus
with no fine-tuning of parameters at all conformal cosmology leads us right 
into the $\bar{\Omega}_M(t_0)=0$, $0 \leq \bar{\Omega}_{\Lambda}(t_0) \leq 1$ 
region favored by the supernovae data. Moreover, with the suppression of 
$\bar{\Omega}_M(t_0)=0$ being achievable without any constraint being put on 
$\rho_M(t_0)$, its value is thus not constrained to be of order the critical 
density $\rho_c=3c^2H^2(t_0)/8\pi G$, with conformal cosmology thus being 
released from the need to contain copious amounts of cosmological dark 
matter.\footnote{Like $\bar{\Omega}_M(t)$, in the conformal case the quantity 
$\Omega_M(t)$ itself starts off being infinite at $t=0$ (no matter what the 
numerical values of the parameters) and finishes up being zero at $t=\infty$. 
Thus, $\Omega_M(t)$ has to pass through one in some epoch without the need for 
any fine tuning (and even has to pass through one quite slowly since the 
cosmology associated with $R(t)$ of Eq. (\ref{36}) is a very slow coasting one) 
while being far from one in other epochs, to thus not constrain the value of 
$\rho_M(t_0)$ at all. Moreover, if dark matter is not invoked, known explicitly 
visibly established matter alone leads to an $\Omega_M(t_0)$ of order $10^{-2}$ 
or so, so $\Omega_M(t_0)$ may not be so close to one today as to require any 
particular explanation in the first place (i.e. the value of $\Omega_M(t_0)$ 
would not be special unless it was actually equal to one to some incredibly high 
degree of accuracy); and indeed, even in an $\Omega_M(t_0) <1$ standard theory, 
ongoing expansion will eventually lead to an $\Omega_M(t)$ which will 
be nowhere near to one, with its current closeness to one then being an 
accidental consequence of the fact that it is this particular epoch in which we 
just happen to be making observations.} Conformal gravity thus not only gets rid 
of the need for galactic dark matter, it eliminates the need for cosmological dark 
matter as well.

With conformal gravity having thus eliminated the need for cosmological dark 
matter, it is of interest to see just exactly how the conformal gravity 
evolution equation of Eq. (\ref{25}) itself actually manages to avoid any 
flatness fine tuning problem.\footnote{The author is indebted to Dr. M. Turner 
for asking a helpful question in this regard.} Thus in the illustrative 
$\lambda=0$ case where the evolution equation is given by 
\begin{equation}
\dot{R}^2(t) +3\dot{R}^2(t)\Omega_M(t)/4\pi S_0^2 L^2_{Pl}=-kc^2
\label{42}
\end{equation}
with solution \cite{Mannheim1998}
\begin{equation}
R^2(t,\alpha=0,k<0)=-2A/\hbar kS_0^2c -kc^2t^2,
\label{43}
\end{equation}
we see the two terms on the left hand side of Eq. (\ref{42}) have radically 
different time behaviors, even as their sum remains constant (=$-kc^2$). 
Specifically, $\dot{R}^2$ begins	at zero and slowly goes to $-kc^2$ at late 
times, while the $3\dot{R}^2(t)\Omega_M(t)/4\pi S^2_0 L^2_{Pl}$ term does the 
precise opposite as it goes to zero from an initial value of $-kc^2$. Moreover, 
while these two terms turn out to be of the same magnitude at some time in the 
early universe, simply because the scale factor $S_0$ is so much larger than 
$L^{-1}_{Pl}$, these two terms are nowhere near the same order of magnitude 
today, and yet their sum remains constant. As such this behavior differs 
radically from that found in the standard model, since there the very fact that 
$L_{Pl}$ is its natural scale forces the analogous two terms (terms which are of 
opposite sign in the standard model) to be of the same order of magnitude at all 
times right up to the present and to thus have to cancel to an extraordinary 
degree. Since such fine-tuning is not required in the conformal case, we see 
that it is the changing of the effective $G$ which explicitly enables us to 
resolve the flatness problem. 

With the general $\alpha \neq 0$ Hubble parameter obeying Eq. (\ref{40b}), we see 
that its current value obeys $-q(t_0)H^2(t_0)=\alpha c^2$, with the current age 
of the universe then being given by $H(t_0)t_0=
arctanh[(-q(t_0))^{1/2}]/(-q(t_0))^{1/2}$. Thus we see that $t_0$ is necessarily 
greater or equal to $1/H(t_0)$ ($t_0=1/H(t_0)$ when $q(t_0)=0$, and 
$t_0=1.25/H(t_0)$ when $q(t_0)=-1/2$). Thus we see that conformal cosmology 
readily resolves \cite{Mannheim1998,Mannheim1999} another problem which has
troubled the standard theory, viz. the universe age problem. Further, in
conformal cosmology the (dimensionless) ratio of the particle horizon size
$d(t)$ to the  spatial radius of curvature $R_{curv}(t)$
(=$(6/R^{(3)})^{1/2}$ where $R^{3}$ is the modulus of the Ricci scalar of
the spatial part of the metric) is given by
\begin{equation}
{d(t) \over R_{curv}(t)}=(-k)^{1/2}c\int_0^t {dt \over R(t)}
=(-k)^{1/2}c\int_0^t {dt \over  [-k(\beta-1)/2\alpha
-k\beta sinh^2 (\alpha^{1/2} ct)/\alpha)]^{1/2}}.
\label{44}
\end{equation}
Thus, in the illustrative $\lambda=0$ case where $q(t_0)=0$ and where 
$T^2_{max}=T^2(t_0)(1-1/\bar{\Omega}_M(t_0))$, Eq. (\ref{44}) evaluates as
\begin{equation}
{d(t) \over R_{curv}(t)}=
log \left[{T_{max} +
(T_{max}^2 -T^2(t))^{1/2} \over T(t)} \right],
\label{45}
\end{equation}
with the horizon size thus being altogether greater than one at recombination. 
Similarly in the $q(t_0)=-1/2$, $T_{max} \gg T_V \gg T(t_0)$ case where the 
current time obeys $sinh^2(\alpha^{1/2}ct_0)=1$ and where $\beta \simeq 1$, we
then find for all times up to recombination that $R(t, \alpha >0, k<0)$ can be 
approximated by $ -k(\beta-1)/2\alpha-k c^2 t^2$, with the recombination time 
horizon size then being found to again be given by Eq. (\ref{45}) (with $T_{max}$ 
being the $\alpha>0$ one this time). Conformal cosmology thus readily resolves 
\cite{Mannheim1998} the horizon problem, and leads to a naturally causally 
connected cosmology.\footnote{For comparison, we recall that in the typical 
$k<0$, $\Lambda=0$ standard gravity case, this same  $d(t)/R_{curv}(t)$ ratio 
is given by $log[(T_{ref}+(T_{ref}^2+T^2(t))^{1/2})/T(t)]$ (where $T^2_{ref}=
T^2(t_0)(1/\Omega_M(t_0)-1)$) and is thus much smaller than one at 
recombination. Thus with the conformal gravity $T_{max}$ being much larger than 
the standard gravity $T_{ref}$ simply because the effective conformal $G$ is so 
much smaller than the standard one, we see that it is the changing of the 
effective $G$ which explicitly enables us to resolve the horizon problem.} Thus, 
to sum up, we see that simply by modifying the effective cosmological $G$, 
conformal gravity is then able to resolve a whole variety of current 
cosmological problems, viz. the flatness, horizon, dark matter, universe age, 
cosmic acceleration and cosmological constant problems, and should thus be seen 
as a potentially viable candidate cosmological theory (for its current overall 
status and for the challenges that it itself still faces see 
\cite{Mannheim1998,Mannheim1999}).

With conformal cosmology being rendered singularity free through the 
negative sign effective Einstein-Hilbert action present in the conformal matter 
action of Eq. (\ref{19}), we see that in conformal gravity it is gravity itself 
which can protect itself from its own singularities. Conformal gravity thus 
provides a possible (though currently far from guaranteed) mechanism by which 
the collapse of a star can still possibly be prevented when all conventional 
mechanisms (such as radiation pressure or Pauli degeneracy) have failed, viz. 
that a scalar field condensate could be generated inside the star (perhaps when 
the star reaches nuclear density) which would then induce some repulsive gravity
(cf. negative energy density\footnote{Such a negative contribution is simply 
not considered in models in which positivity of the energy density is assumed, 
and even while the energy density of a standard kinematic perfect fluid is 
indeed positive, we thus see that the extension of such positivity to the entire 
energy-momentum tensor which serves as the source of gravity is actually 
questionable in general (even if absent in the flat space limit, non-inertial 
explicitly curvature dependent terms are not forbidden in the curved space case), 
as would then be the gravitational collapse theorems which rely on such 
positivity and assume that no negative component is ever generated during any 
such collapse.}) and arrest the collapse (to then either generate a rebound or 
produce a stable configuration with radius greater than the Schwarzschild radius 
of the star).\footnote{While recent data \cite{Richstone1998} on the velocities 
of stars in inner galactic regions have indicated the presence of large (black 
hole candidate) mass concentrations at the centers of galaxies such as M87 and 
the Milky Way, those data are currently unable to ascertain whether any such 
large mass is actually confined to a radius smaller than its Schwarzschild 
radius or determine whether any event horizon has actually been formed (at 400 
km s$^{-1}$ or so the measured velocities, while large by galactic standards, 
are still well below the velocity of light).} Thus with gravity being able to 
generate its own repulsion, the whole issue of gravitational singularities needs 
to be reconsidered, with the standard strong gravity picture possibly being 
another piece of the standard weak gravity intuition whose generalization might 
be unreliable.

Having thus presented the cosmological case for a negative spatial curvature 
universe, we now present some additional, quite direct observational evidence 
in its support, evidence from an at first somewhat unlikely source, namely the 
systematics of galactic rotation curves. While these curves provided the first 
clear evidence of the need in standard gravity for dark matter, beyond the fact 
that these curves show that there actually is a departure from the luminous 
Newtonian expectation in the outskirts of spiral galaxies, explicit study 
\cite{Mannheim1997,Mannheim1996} of the systematics of such departure has 
revealed the presence of an apparent cosmological imprint on the data, with it 
being cosmology itself which will actually enable us to eliminate 
the need for any galactic dark matter at all. Indeed, with the potentials
associated with static sources in conformal gravity being ones which
actually grow with distance according to Eq. (\ref{7}), the very fact that
they do so entails that in calculating the motions of individual particles
within a given galaxy, one is now  no longer able to ignore the contributions
of the potentials due to distant matter sources outside of that 
galaxy. Thus in going to a higher order theory such as conformal gravity,
we immediately transit into a world where we have to consider effects due to
matter not only inside but also outside of individual systems, and thus we
are led to look for both local and global imprints on galactic rotation
curve data, this being a quite radical (and quite Machian) conceptual
departure from the standard second order, purely local Newtonian world 
view.\footnote{With it being only for spherically symmetrically distributed 
matter with $1/r$ potentials that exterior sources decouple locally, the 
very detection of any global cosmological imprint within galaxies would then 
argue against gravitational potentials being of a pure $1/r$ form.} 

To isolate such possible global imprints, it is instructive \cite{Mannheim1997} 
to look at the centripetal accelerations of the data points farthest from 
the centers of individual galaxies. In particular, for a set of 11 particular
galaxies whose rotation  curves are regarded as being particularly
characteristic of the pattern of  deviation from the luminous Newtonian
expectation that has so far been obtained,  it was found that the farthest
centripetal accelerations in these galaxies could  all be parameterized by
the universal three component relation
\begin{equation}
(v^2/R)_{last}=\gamma_0c^2/2+\gamma^{*}N^{*}c^2/2 +\beta^{*}N^{*}c^2/R^2
\label{46}
\end{equation}
where $\gamma_0=3.06\times 10^{-30}$ cm$^{-1}$, 
$\gamma^{*}=5.42\times 10^{-41}$ cm$^{-1}$, $\beta^{*}=1.48\times 10^5$ cm, and 
where $N^{*}$ is the total amount of visible matter (in solar mass units) in 
each galaxy. Since the luminous Newtonian contribution is decidedly non-leading
at the outskirts of galaxies, we thus uncover the existence of two linear 
potential terms which together account for the entire measured departure 
from the luminous Newtonian expectation,\footnote{With the luminous Newtonian 
contribution falling with distance and with the rotation curves of the prominent 
bright spirals being flat, the departure from the luminous Newtonian expectation
must itself thus be growing with distance, and according to Eq. (\ref{46}) even 
be growing universally in fact.} with one of these two terms depending on
the number, $N^{*}$, of stars within each given galaxy, and with the other, the 
$\gamma_0c^2/2$ term, not being dependent  
on the mass content of the individual galaxies at all. Moreover, since 
numerically $\gamma_0$ is found to have a magnitude of order the inverse Hubble 
radius, we can thus anticipate that it must represent a universal global effect 
generated by the matter outside of each galaxy (viz. the rest of the matter in 
the universe), and thus not be associated with any local dynamics within 
individual galaxies at all.

As regards the $N^{*}$ dependent contribution of the matter within the
individual galaxies, the integration of the non-relativistic stellar
potentials
$V^{*}(r)=-\beta^{*}c^2/r+\gamma^{*}c^2r/2$ over an infinitesimally thin
galactic optical disk with luminous surface  matter distribution
$\Sigma(R)=\Sigma_0$exp$(-R/R_0)$ and total number of  stars $N^{*}=2 \pi
\Sigma_0 R_0^2$ yields \cite{Mannheim1997} the centripetal acceleration
\begin{equation}
v^2/R=g_{gal}^{lum}=g_{\beta}^{lum}+g_{\gamma}^{lum} 
\label{47}
\end{equation}
where
\begin{equation}
g_{\beta}^{lum}=
(N^{*}\beta^{*} c^2 r/2R_0^3)[I_0( r/2R_0)K_0( r/2R_0)-
I_1( r/2R_0)K_1(r/2R_0)]
\label{48}
\end{equation}
and where
\begin{equation}
g_{\gamma}^{lum}=(N^{*}\gamma^{*} c^2r/2R_0)I_1( r/2R_0)K_1( r/2R_0),
\label{49}
\end{equation}
to thus yield a net acceleration which behaves asymptotically as
$(v^2/R)_{last}=\gamma^{*}N^{*}c^2/2 +\beta^{*}N^{*}c^2/R^2$. The conformal
gravity local galactic potentials associated with the matter within any
given galaxy thus nicely generate the
$N^{*}$ dependent terms  exhibited in Eq. (\ref{46}). 

As regards the remaining $N^{*}$ independent $\gamma_0c^2/2$ term, in order to
be able to identify it as being of cosmological origin, we need to rewrite
the comoving Hubble flow of the rest of the universe in the rest frame 
coordinate system of any given galaxy of interest. To determine just how
a comoving geometry might look in a Schwarzschild coordinate system, we note
\cite{Mannheim1989} that the general coordinate transformation
\begin{equation} 
r=\rho/(1-\gamma_0 \rho/4)^2,~~~ t = \int d\sigma / R(\sigma)
\label{50}
\end{equation} 
effects the metric transformation 
\begin{eqnarray}
d\tau^2=(1+\gamma_0 r)c^2dt^2-{dr^2 \over (1+\gamma_0 r)}-r^2d\Omega
\rightarrow
\nonumber \\
{(1+\rho\gamma_0/4)^2 \over R^2(\sigma)(1-\rho\gamma_0/4)^2}
\left(c^2 d\sigma^2 - {R^2(\sigma) (d\rho^2 + \rho^2 d\Omega)\over
(1-\rho^2\gamma_0^2/16)^2}
 \right). 
\label{51}
\end{eqnarray}
Thus with metrics conformal to a Robertson-Walker one also being allowed
cosmological solutions in a conformal invariant theory, we see that in
conformal gravity a static, Schwarzschild coordinate linear potential metric
is coordinate equivalent to conformal cosmologies in which the spatial
curvature $k=-\gamma_0^2/4$ is expressly negative,\footnote{Positive $k$
would lead to a complex $\gamma_0$ in Eq. (\ref{51}), with it not being
possible for the  topologically open $d\tau^2=(1+\gamma_0 r)c^2dt^2-dr^2
/(1+\gamma_0 r) -r^2d\Omega$ metric to ever be equivalent to anything other
than other  topologically open ones.} with a universal linear potential thus
being the local manifestation of global negative curvature, and with the
local $\gamma_0$ and the global $k$ thus having a common connection. As
such, this connection is actually geometrically quite natural, since not only
does negative $k$ lead to repulsion, but, as had been noted earlier, so does
positive $\gamma_0$. However, while a positive $\gamma_0 r$ metric term does
indeed lead to gravitational deflection of light, nonetheless, for
non-relativistic systems this same positive $\gamma_0 r$ term acts
attractively, with cosmological negative spatial curvature thus generating
an attractive gravitational effect for non-relativistic motions within
galaxies, an effect which the standard theory in essence tries to simulate
locally by the introduction of local galactic dark matter. Now while we
had initially identified the static $\gamma_0$ term of Eq. (\ref{46}) as being
of cosmological origin since its phenomenologically measured value was found
to be of order the inverse of the Hubble radius, such an identification could
at best have only been heuristic since the Hubble parameter itself is a time
dependent one which varies from one epoch to the next. However, cosmology
actually possesses a second scale beyond that associated with its expansion
rate, namely that associated with its spatial curvature, with it being this
latter, epoch independent one, with which $\gamma_0$ is then nicely
identified, with the $\gamma_0 r$ term thus serving as a time independent 
universal potential term no matter what the epoch.  

Further, for galaxies which have no peculiar velocities with respect to the
Hubble flow, i.e. for any galaxy whose center can precisely serve as the
coordinate origin for  the radial coordinate $r$ in the transformation of Eq.
(\ref{50}), the  metrics of Eqs. (\ref{47}) and (\ref{51}) can simply be
added in the weak gravity limit, to yield as the net weak gravity centripetal
acceleration 
\begin{equation}
v^2/R=g_{tot}=g_{\beta}^{lum}+g_{\gamma}^{lum}+\gamma_0 c^2/2,
\label{52}
\end{equation}
an acceleration whose asymptotic limit precisely yields Eq. (\ref{46}) for
$(v^2/R)_{last}$. Moreover, not only does Eq. (\ref{52}) yield this requisite
asymptotic formula, its use in the sub-asymptotic region as well is then found
to provide acceptable parameter free (and thus dark matter free) fitting 
\cite{Mannheim1997,Mannheim1996} to all of the (in excess of 250) data points in 
the 11 galaxy 
sample, with Eq. (\ref{52}) thus capturing the essence of the data.\footnote{For 
comparison, standard dark matter halo fitting uses two free parameters per halo 
and thus no less than 22 additional free parameters for the same 11 galaxy 
sample.} Thus we identify an explicit imprint of cosmology on galactic rotation 
curves, recognize that it is its neglect which may have led to the need for dark 
matter, and for our purposes here confirm that $k$ is indeed negative, just as 
had been required in the cosmological study which we presented above, with
the phenomenological formula of  Eq. (\ref{46}) actually providing an explicit
measurement of the curvature of the universe which test particles sample as
they orbit in galaxies.

Moreover, not only does Eq. (\ref{52}) provide for an acceptable accounting of 
galactic rotation curve data, the magnitude obtained for the stellar $\gamma^{*}$ 
is found to be so small (of order $10^{-41}$ cm$^{-1}$) that the linear potential 
term then makes a completely negligible contribution on solar system 
distance scales, with the metric of Eq. (\ref{7}) thus reducing to the standard 
Schwarzschild one within the solar system. The strength of the linear potential 
terms in Eq. (\ref{52}) thus serve as the scale which is to parameterize 
departures from the luminous Newtonian expectation (for the bright spirals 
$\gamma^{*}N^{*}$ and $\gamma_0$ are of the same order of magnitude), with such a 
scale nicely explaining why no dark matter is needed on solar system 
distance scales, with the solar system simply being too small to be sensitive to 
any cosmologically relevant scale. With the linear potential term first becoming  
competitive with the Newtonian one on none other than galactic distance scales, 
we thus explain not only why solar system physics is unaffected by conformal 
gravity, but also, we identify at exactly what point departures from the luminous 
Newtonian contribution are to first set in. The (negative) spatial curvature
of the  universe thus sets the scale at which standard gravity needs to
introduce dark  matter in order to avoid failing to fit data. 

Our uncovering of a universal acceleration in conformal gravity immediately 
recalls the presence of a similar one in the MOND theory \cite{Milgrom1983}, one 
also of a cosmologically significant magnitude. Specifically, Milgrom had 
suggested that if a universal acceleration $a_0$ did exist, then Newton's law of 
gravity could possibly be phenomenologically modified into a form such as 
$v^2/R=\nu(a_0/g_N)g_N$ where $g_N=g_{\beta}^{lum}$ in the galactic case.
The  candidate functional form $\nu(x)=(1/2+(4x^2+1)^{1/2}/2)^{1/2}$ would
then yield
\begin{equation}
v^2/R=g_N \{1/2+ (g_N^2+4a_0^2)^{1/2} / 2g_N \}^{1/2},
\label{53}
\end{equation}
an expression which, despite the absence so far of any deeper underlying 
theory, is nonetheless found to perform extremely well phenomenologically. 
While Eqs. (\ref{52}) and (\ref{53}) have different underlying 
motivations, it is of interest to note that Eq. (\ref{52}) would in fact fall 
into the general MOND approach if the MOND formula were to be reinterpreted as
\begin{equation}
v^2/R=\nu(\gamma_0c^2/2g_{loc})g_{loc}
\label{54}
\end{equation}
where $g_{loc}$ is the entire local luminous galactic contribution 
$g_{gal}^{lum}$ given in Eq. (\ref{47}), and if the function $\nu(x)$ were to 
instead take the form $\nu(x)= 1+x$. Conformal gravity thus not only
provides  a rationale for why there is in fact a universal acceleration in
the first place  (something simply assumed in MOND), but also it even yields
an explicit form for the function $\nu(x)$, albeit not the one commonly
utilized in the standard  MOND studies (where $g_{loc}$ is taken to be the
purely Newtonian
$g_N$).  However, despite such differences, both theories have in common the
recognition  that there is a universal scale which is to parameterize
departures from the  standard luminous Newtonian expectation, and that its
magnitude is a cosmologically significant one. 

Further support for the existence of such a scale has been presented by McGaugh        
\cite {McGaugh1998} in a study of the behavior of the quantity
$M_{dyn}(R)/M_{lum}(R)$ as a function of the measured orbital acceleration 
$v^2(R)/R$ at points $R$ within galaxies. ($M_{dyn}(R)=Rv^2(R)/G$ is the
amount of matter interior to $R$ as would be required by Newtonian gravity, 
while $M_{lum}(R)$ is the amount of
luminous matter detected in the same region). In this study McGaugh found
that mass discrepancies (viz $M_{dyn}(R)/M_{lum}(R)>1$) systematically
occurred  in galaxies whenever the measured $v^2(R)/R$ fell below a universal
value of 
$10^{-8}$ cm s$^{-2}$ or so, a value which is immediately recognized as being 
close to the values of the $a_0$ and $\gamma_0^2c^2/2$ acceleration parameters 
which were respectively phenomenologically obtained in the MOND and conformal 
gravity theories. While MOND and conformal gravity might differ as to how the 
centripetal accelerations should behave in regions where there are measured mass 
discrepancies, both theories (and the data) thus agree that there is a
universal  scale which determines when such discrepancies should first set in.

Hence, independent of the merits of alternate theories such as conformal gravity 
or MOND themselves, it would appear that the data possess a cosmological imprint,
an imprint which heralds when dark matter is first needed in the standard
theory, with the very existence of such an imprint enabling Eqs. (\ref{52})
and (\ref{53}) to organize the data in an extremely economical fashion. Thus
even if one does not want to  contemplate going beyond standard gravity, the
galactic data themselves seem to  be insisting that dark matter theories
should be parameter free (i.e. that they should be formulatable without the
(extravagant) need for two free parameters per  galactic halo), and that
there is a cosmological imprint in the data which dark  matter theories must
be able to produce. Moreover, apart from the fact that  dynamical dark matter
models have not yet produced such a scale (say by studying  the growth of
galaxy fluctuations in cosmology), it would appear difficult for  them to
ever be able to do so in the standard cold dark matter flat inflationary 
universe cosmological model, since simply by virtue of being flat such a 
cosmology then lacks the one key ingredient which leads to a universal
galactic  scale in the conformal theory, namely a non-zero spatial curvature,
with the  standard flat classical cosmological model simply not possessing
any intrinsic  such universal scale at all.\footnote{See \cite{Periwal1999}
for an interesting  attempt to generate such a scale quantum-mechanically as
a renormalization group  correlation length associated with a quantum gravity
fixed point, a point at  which scale invariance (such as that in conformal
gravity) is realized via  anomalous dimensions.} Given the above
observational support for the existence  of such a scale, galactic data thus
seem to be supporting the notion that the universe has a non-zero spatial
curvature, a non-trivial curvature which  conformal gravity (and for the
moment only conformal gravity apparently) can readily and  naturally produce,
a curvature which releases gravity from having to always be attractive.

Thus to conclude, we believe that it is not yet justified to assert that gravity 
is always attractive, and that in fact a repulsive cosmological component 
immediately allows one to resolve a whole host of problems which currently 
beset the standard theory. And even if conformal gravity itself should not
turn  out to be the correct extrapolation of the standard solar system wisdom
(as a  cosmology conformal gravity is not without challenges of its own 
\cite{Mannheim1998,Mannheim1999}), that would in no way constitute evidence in
favor of the correctness of the standard extrapolation. Since our study of
conformal  gravity has shown that the problems with which the standard theory
is currently  afflicted are not in fact generic to cosmology, their very
existence could be a  warning that the extrapolation of standard gravity
beyond the confines of the  solar system might be a lot less reliable than is
commonly believed, with gravity not always being as attractive as Newton
initially took it to be.
   
\section{Author's Note}
During the last few years I have written articles 
\cite{Mannheim1994a,Mannheim1996} on gravitational theory for 
special issues of Foundations of Physics in honor of my longtime colleagues 
and friends Fritz Rohrlich and Larry Horwitz. It is a great pleasure for me to 
dedicate this third article in that series to another equally close colleague 
and friend Kurt Haller. This work has been supported in part by the Department 
of Energy under grant No. DE-FG02-92ER4071400.

\end{document}